\documentclass{article}

\PassOptionsToPackage{sort&compress, numbers, square}{natbib}

\usepackage[preprint]{neurips_2025}

\usepackage[utf8]{inputenc} 
\usepackage[T1]{fontenc}    
\usepackage{hyperref}       
\usepackage{url}            
\usepackage{booktabs}       
\usepackage{amsfonts}       
\usepackage{nicefrac}       
\usepackage{microtype}      
\usepackage{xcolor}         
\usepackage{tabularx}

\usepackage{amsmath}
\usepackage{amssymb}
\usepackage{multirow}
\usepackage{graphicx}

\title{Self-DANA:\\A Resource-Efficient Channel-Adaptive Self-Supervised Approach for\\ECG Foundation Models}

\author{\\\textbf{Giuliana~Monachino}\textsuperscript{1,2*}, \textbf{Nicolò~La Porta}\textsuperscript{3,4}, \textbf{Beatrice~Zanchi} \textsuperscript{1,5}, \textbf{Luigi~Fiorillo}\textsuperscript{1,6}, \\ \\ \textbf{Alvise~Dei Rossi}\textsuperscript{1,4}, \textbf{Georgiy~Farina}\textsuperscript{1}, \textbf{Francesca D.~Faraci}\textsuperscript{1}}

\begin{document}

\maketitle
\vspace{-0.75cm}
\begin{center}
\begin{it}
    \textsuperscript{1} Institute of Digital Technologies for Personalized Healthcare (MeDiTech),\\Department of Innovative Technologies, University of Applied Science and Art of Southern Switzerland (SUPSI), Lugano, 6962, Switzerland\\
    \textsuperscript{2} Institute of Informatics, University of Bern, Bern, 3012, Switzerland\\
    \textsuperscript{3} Institute of Information Systems and Networking (ISIN), Department of Innovative Technologies, University of Applied Science and Art of Southern Switzerland (SUPSI), Lugano, 6962, Switzerland\\
    \textsuperscript{4} Faculty of Informatics, Università della Svizzera Italiana (USI), Lugano, 6900, Switzerland\\
    \textsuperscript{5} Department of Quantitative Biomedicine, University of Zurich, Zurich, 8091, Switzerland\\
    \textsuperscript{6} Neurology Department, Neurocenter of Southern Switzerland,\\Ente Ospedaliero Cantonale (EOC), Lugano, 6962, Switzerland\\
\end{it}
\vspace{0.5cm}
\textsuperscript{*}Corresponding author: \texttt{giuliana.monachino@supsi.ch}

\end{center}
\vspace{1cm}

\begin{abstract}
Foundation Models (FMs) are large-scale machine learning models trained on extensive, diverse datasets that can be adapted to a wide range of downstream tasks with minimal fine-tuning. In the last two years, interest in FMs has also grown for applications in the cardiological field to analyze the electrocardiogram (ECG) signals. 
One of the key properties of FMs is their transferability to a wide range of downstream scenarios. With the spread of wearable and portable devices, keen interest in learning from reduced-channel configurations has arisen. However, the adaptation of ECG FMs to downstream scenarios with fewer available channels still has to be properly investigated.
In this work, we propose Self-DANA, a novel, easy-to-integrate solution that makes self-supervised architectures adaptable to a reduced number of input channels, ensuring resource efficiency and high performance. We also introduce Random Lead Selection, a novel augmentation technique to pre-train models in a more robust and channel-agnostic way.
Our experimental results on five reduced-channel configurations demonstrate that Self-DANA significantly enhances resource efficiency while reaching state-of-the-art performance. It requires up to 69.3\% less peak CPU memory, 34.4\% less peak GPU memory, about 17\% less average epoch CPU time, and about 24\% less average epoch GPU time.
\end{abstract}

\clearpage
\section{Introduction}
\label{sec:introduction}
Foundation Models (FM), first formally defined in \cite{Bommasani2021_FMdef}, are large-scale machine learning models "\textit{trained on broad data (generally using self-supervision at scale) that can be adapted
(e.g., fine-tuned) to a wide range of downstream tasks}". 
A pivotal role in the development and diffusion of FMs was played by advances in Self-Supervised Learning (SSL), a training paradigm where models learn useful general representations from unlabeled data by solving auxiliary (pretext) tasks.
The success of models like BERT \cite{Devlin2019_BERT}, GPT-3 \cite{Brown2020_GPT}, and CLIP \cite{Radford2021_CLIP}, demonstrated the power of pre-training on general data followed by minimal task-specific adaptation. 
First introduced in the natural language processing field \cite{Brown2020_GPT, Devlin2019_BERT}, then FMs gained traction also in computer vision \cite{Radford2021_CLIP}, audio processing \cite{Baevski2020_wav2vec2, Hsu2021_HuBERT}, and recently reached the biosignals domain too \cite{Thapa2025_sleepFM, Vaid2023_HeartBEiT}.\\
This work primarily focuses on the Electrocardiogram (ECG), a biomedical signal that describes the heart's electrical activity, revealing information about its rhythm, morphology, and potential abnormalities. Thorough and timely analysis of ECG signals is essential for accurate diagnosis and effective treatment. Machine Learning (ML) and Deep Learning (DL) techniques have long been explored and applied in automated ECG signal analysis, achieving remarkable performance by extracting meaningful patterns and features from large datasets.\\
In recent years, a growing interest in SSL methods for ECG analysis \cite{DelPup2023_review_SSL_biosig, Han2024_reviewFM} has emerged, leading to the advent of the first ECG FMs \cite{McKeen2024_ECGFM, Abbaspourazad2023, Yu2024, Mathew2024, Vaid2023_HeartBEiT}.
Most of them rely on Self-Supervised Contrastive Learning (SSCL), which is based on learning meaningful representations by solving a contrastive pretext task: training the model to differentiate between positive pairs (i.e., augmented versions of the same input) and negative pairs (i.e., augmented versions of different inputs).

One of the key properties of FMs is transferability, i.e., the ability to generalize to a wide range of downstream scenarios with zero or minimal fine-tuning. Their SSL foundations allow them to learn from the structure of raw data itself, requiring minimal labeled data for adaptation, and transferring effectively across tasks.
Their strong generalization significantly lowers the need for extensive fine-tuning, enabling the achievement of improved performance with limited data and computational resources.
A key yet underexplored aspect in the ECG literature is FMs' adaptability from the standard 12-lead configuration to downstream scenarios involving a reduced number of channels (or \textit{leads}), i.e., \textit{reduced-channel configurations} or \textit{reduced-lead configurations}.
This aspect is particularly crucial in the wearables or portable devices domain, which often operates with a limited number of sensors or channels due to size, power consumption, and user comfort constraints. Indeed, although the 12-lead ECG remains the clinical standard to diagnose many cardiac conditions, several portable ECG devices use fewer leads to enable continuous monitoring \cite{bouzid2022remote}. Common chest straps exploit 1-3 leads, some handheld devices allow 6 leads recording, while smartwatches record only a single-lead ECG (lead I).\\
Most of the literature related to ECG FMs keeps the same channel configuration (usually the standard 12-lead) for pre-train and fine-tune/test, being forced to pre-train ad-hoc models with fewer leads in case of reduced-lead downstream tasks (e.g. \cite{Abbaspourazad2023}).
Some other techniques rely on the combination of 12 channels, so this approach would not even be possible \cite{Gopal2021_3KG, Liu2023_JCDCL}.
These approaches conflict with the core idea of FM and the principle of transferability and generalization. Indeed, the best solution would be to have a single FM able to adapt to any downstream scenario with minimal fine-tuning.
To date, the only channel-agnostic ECG FM present in the literature is \cite{McKeen2024_ECGFM}, which adopts the technique proposed by \cite{Oh2022_RLM}, including Random Lead Masking (RLM) and zero-padding (see Section~\ref{sec:related_works} for further details).
However, the use of zero-padding makes this technique computationally inefficient, especially when only a few leads are available for the downstream task.

In this work, we propose Self-DANA, a new, easy-to-integrate solution to make self-supervised architectures adaptable to a reduced number of input channels while ensuring resource efficiency and high performance. It combines a dimension-adaptive architecture with an ad-hoc augmentation for self-supervised contrastive learning pre-training. This combination allows to obtain a FM adaptable and robust to diverse channel availability, while keeping the fine-tuning phase efficient from a memory and computation time standpoint.

Our main contributions are the following:
\begin{itemize}
    \item \textbf{Dimension Adaptive Pooling (DAP) layer adoption.} We suggest the adoption of a DAP layer for building dimension-adaptive backbone architectures to enable FMs to adapt to reduced-lead configurations in downstream tasks. This is an easy-to-integrate and resource-efficient alternative to zero-padding.
    \item \textbf{Random Lead Selection (RLS).} We introduce RLS, a new contrastive-learning augmentation that enables the model to learn more robust ECG representations.
    \item \textbf{Self-DANA.} We propose Self-DANA, a simple approach that combines the DAP layer with RLS, to enhance its potential, acting as an ad-hoc augmentation for channel-adaptive SSCL-based algorithms.
    \item \textbf{Evaluation on reduced-channel out-of-domain scenarios.} We evaluate our approach on an out-of-domain dataset, on five different reduced-channel configurations.
\end{itemize}

\section{Related works}
\label{sec:related_works}

\paragraph{Random Lead Masking for lead-agnostic ECG FM} \cite{McKeen2024_ECGFM} proposed a channel-agnostic FM, building upon the local-global lead-agnostic approach introduced and tested by \cite{Oh2022_RLM} to develop a larger open foundation model named \textit{ECG-FM}.
\cite{Oh2022_RLM} introduced Random Lead Masking (RLM), an ECG-specific augmentation method to improve the model's robustness to a reduced-lead configuration. During pre-training, it masks 6 out of 12 randomly chosen channels. 
In the fine-tuning phase, if only a subset of channels is available for the downstream task, the missing leads are filled with zeros (\textit{zero-padding}) to maintain the expected 12-channel format.
They evaluated the approach on two downstream tasks (diagnosis classification and patient identification), exploiting in-domain datasets and different lead configurations (12-lead, 6-lead (I, II, III, aVF, aVL,
aVR), 3-lead (I, II, V2), 2-lead (I, II), and 1-lead (I)). Their results demonstrated that RLM augmentation enables the model to learn from diverse lead combinations, leading to improved performance compared to the same method without RLM. The main drawback of this approach is its poor resource efficiency. Indeed, even when only a single lead is available for a downstream task, the model still needs to be fine-tuned by processing inputs padded to the 12-channel format used during pre-training. This implies that 11 out of the 12 channels, containing only zeros, would consume significant memory without contributing meaningful information and could potentially introduce unwanted bias.
\paragraph{Dimension Adaptive Neural Architecture}
\cite{Malekzadeh2020_DANA} proposed a combined use of a Dimension Adaptive Pooling (DAP) layer with a Dimension Adaptive Training (DAT) procedure to make deep neural networks robust to changes in sensor availability and sampling rate, resulting in a Dimension Adaptive Neural Architecture (DANA).
The DAP layer is an average pooling layer with a variable kernel size that can be adapted to match any input dimension to a fixed output dimension. 
Since Convolutional Neural Networks (CNN) are inherently adaptive to variable input data dimensions, the authors propose the DAP layer for architectures composed of a first convolutional part and a second non-adaptive part (e.g., recurrent or feedforward neural network). They suggest placing the adaptive layer between the last convolutional layer and the first feedforward/recurrent layer.
In contrast, the DAT procedure involves training the model using inputs with randomly selected dimensions and sensor orders, enhancing its robustness to varying input availability during the inference phase.\\
With the DANA approach, they demonstrated the ability to train a single adaptive model with high classification performance across diverse settings, without the need to train a different classifier for any possible setting.
\paragraph{Other channel-agnostic related works} \cite{Li2024_ECGFounder} proposed \textit{ECGFounder}, a supervised foundation model pre-trained on the largest open-access ECG dataset. 
Besides the standard 12-lead model, the authors designed a single-lead model for wearables. It exploits a special training strategy based on lead augmentation to learn information from lead I, as if it had been extracted from six derivations. 
Despite the validity of the work, we cannot consider this approach as adaptable to any reduced-lead configuration since it is built only for a specific one (lead I configuration).
\cite{Yang2023_BIOT} and \cite{Thapa2025_sleepFM} are also relevant to our work, despite not focusing on FMs for ECG signals.
\cite{Yang2023_BIOT} presents a technique to overcome missing channels, mismatched sample lengths, and missing values issues in biosignals. They propose to tokenize and embed each channel separately and then flatten them into a unique long "sentence", including also a channel embedding. 
Their approach proved to be more robust to missing channels than other baseline models, but they tested this aspect only in a supervised context and with electroencephalogram (EEG) datasets by randomly dropping up to 25\% of the channels. 
\cite{Thapa2025_sleepFM}, instead, proposes a multi-modal FM for sleep analysis from polysomnography, including a channel-agnostic approach to handle channel variability across datasets. In their architecture, they include a transformer layer to compute attention scores for each channel and output a single embedding per time segment by averaging over the channel dimension.
However, both \cite{Yang2023_BIOT} and \cite{Thapa2025_sleepFM} did not investigate the generalizability of models pre-trained on 12-lead ECGs to reduced-channel downstream scenarios. Hence, we cannot state that these approaches are robust to reduced-channel configurations in such conditions.
Finally, \textit{PhysioNet/Computing in Cardiology Challenge 2021} \cite{cinc21_1, cinc21_2} should be mentioned for the effort in promoting research in channel-agnostic learning. However, the focus of the related works was to solve a specific task with reduced-lead configurations, mainly with supervised approaches, rather than to propose a channel-agnostic FM.

\section{Methods}
\label{sec:methods}
In this work, we propose Self-DANA, a self-supervised approach comprising a dimension-adaptive and resource-efficient architecture (DAP-layer) combined with a novel ad-hoc contrastive-learning augmentation technique (RLS).\\

\subsection{Framework and architecture}
\label{sec:architecture}
\paragraph{SimCLR} 
As a self-supervised learning framework, we adopted SimCLR for our experiment. It was originally designed for computer vision tasks and subsequently adapted to biosignals, laying the foundations for most current SSL approaches for ECG signals \cite{DelPup2023_review_SSL_biosig}. 
Based on the contrastive learning principle, its first steps consist of generating augmented views of the same input, namely \textit{positive pairs}, and encoding them with a shared-weights neural network. Finally, the model is trained to maximize the similarity between positive pairs while minimizing similarity with other samples, namely \textit{negative pairs}, exploiting the Normalized Temperature-scaled Cross Entropy Loss (\textit{NT-Xent loss}) function \cite{Chen2020_simclr}.
The shared-weights neural network comprises an encoder (the \textit{backbone}) and a \textit{projection head}. In our work, we generated the positive pairs with three different sequences of \textit{augmentations} depending on the experiment (see Section~\ref{sec:augmentations} and ~\ref{sec:experiments}).

\paragraph{Dimension-adaptive architecture with DAP layer}
As a backbone, we propose a memory-efficient and dimension-adaptive variant of the architecture exploited in \cite{Oh2022_RLM} inspired by \cite{Baevski2020_wav2vec2}, which integrates a convolutional feature encoder for latent representation extraction, with a transformer encoder to capture long-range contextual dependencies/representations. The transformer encoder comprises a convolutional positional encoder and a sequence of 12 transformer-encoder blocks, each with 12 self-attention heads.
While the transformer encoder exploited in our work is identical to the one used in \cite{Oh2022_RLM}, we introduce a different feature encoder architecture.
Our idea is to make it independent from the number of input channels by placing a DAP layer on top. It reduces any input dimension \((C, T)\) to \((1, 156)\) by applying an average pooling, to match the input dimension required by the subsequent layer. Moreover, we also replaced the original four 1D-convolutional layers with four 2D-convolutional layers with kernel \((1,k)\) and stride \((1,s)\), to apply only temporal convolution, thus ensuring channel independence. 
The backbone is followed by a projection head consisting of a single linear layer, as in \cite{Oh2022_RLM}. Since it was not needed with the SimCLR framework, the masking and quantization operations present in \cite{Oh2022_RLM} have been omitted. 
Except for the newly proposed convolutional feature encoder, we kept the same hyperparameters as in the original work.\\ 
The introduction of the DAP layer allows the proposed feature encoder to considerably reduce resource usage in downstream tasks, particularly when only a few channels are available. Moreover, by applying convolution independently to each channel along the temporal axis, we exploit 5.6K fewer parameters than the original architecture.\\
For the downstream task, a \textit{classification head}, consisting of a fully connected layer with sigmoid activation, is added on top of the backbone. All the architecture details are reported in the Appendix.

\subsection{Contrastive learning augmentations}
\label{sec:augmentations}
\paragraph{Base augmentations}
As a baseline, we selected a set of five basic augmentation techniques usually exploited in the literature for ECG signals and suitable for our context: amplitude scaling, Gaussian noise, crop and resize, time masking, and time warping. Specifically, to encourage the model to learn more robust representations, we create positive pairs by applying first amplitude scaling, followed by one augmentation randomly chosen from the remaining four. For simplicity, in this work, we refer to this sequence of augmentations as \textit{base augmentations}.
More detailed information regarding their implementation and parameters is reported in the Appendix. 

\paragraph{Random Lead Masking (RLM)} Originally introduced in \cite{Oh2022_RLM}, RLM consists of randomly masking a subset of channels by setting their values to 0, while keeping the rest unaltered.
We randomly choose, with uniform probability, both the type (any of the 12 standard ECG leads) and number (between 0 and 11) of channels to mask. 

\paragraph{Random Lead Selection (RLS)}
Our proposed augmentation generates positive pairs by randomly selecting only a subset of the input channels. To ensure dimensional match, this augmentation must be used either with architectures inherently adaptive to variable input data dimensions, like CNNs, or in combination with a DAP layer, like in our experiments.
We randomly choose, with uniform probability, both the type (any of the 12 standard ECG leads) and number (between 1 and 12) of channels to retain. 
 
\subsection{Self-DANA}
The approach proposed in this work extends and optimizes the idea of DANA \cite{Malekzadeh2020_DANA} to SSL frameworks. As in the original formulation, the first component is the introduction of the DAP layer, to make the architecture adaptive to variable input dimension (see Section~\ref{sec:architecture}). In the case of SSL, it also makes pre-trained models adaptable to any channel subset configuration required in the downstream task. In contrast to zero-padding, this technique uses only the available channels, avoiding including additional values that consume memory without contributing meaningful information. The second element of DANA is the dimension adaptive training. We proposed a self-supervised version of this idea, which consists of using our new ad-hoc augmentation, the RLS (see Section~\ref{sec:augmentations}), as (additional) augmentation for the contrastive pretext task. By exploiting the contrastive-learning principle, it encourages the model to generalize across diverse channel combinations, further enhancing the potential of the DAP layer. 

\subsection{Datasets}
\label{sec:datasets}
We pre-trained all the models on a large collection of 12-lead ECG datasets. We then fine-tuned and tested the models on a \textit{out-of-domain} downstream dataset (i.e., never seen during pre-training). The main characteristics of these datasets are reported below, while further details can be found in the Appendix. 
All datasets have been prepared using the same procedure: resampling at 500 Hz, segmentation into non-overlapping 5s windows, and filtering (see Appendix). 

\paragraph{Pre-training}
The collection of data exploited for pre-training is made up of seven open-access datasets: \textit{Code-15\%} \cite{code_train_paper, code_training_data}, \textit{Off-test} \cite{off_test_paper, off_test_data} and five datasets from \textit{PhysioNet/Computing in Cardiology Challenge 2021} \cite{cinc21_1, cinc21_2, PhysioNet, cinc21_data}: (\textit{Ningbo} \cite{ningbo_paper}, \textit{Chapman-Shaoxing} \cite{chapman-shaoxing_paper}, \textit{INCART}, \textit{CPSC} and \textit{CPSC-extra} \cite{cpsc_paper}. This amounts to a total of 406'117 recordings from 295'245 different subjects. After segmentation, a total of 855'424 5-second windows were obtained. All 12 available leads have been used in the pre-training phase. The resulting data collection has been split on a subject basis into training and validation sets according to an 80:20 ratio and source dataset stratification. 

\paragraph{Fine-tuning} 
The experiments comparison has been conducted on the out-of-domain dataset \textit{The Georgia 12-lead ECG Challenge Database} (\textit{Georgia}) \cite{cinc21_1, cinc21_2, PhysioNet, cinc21_data}. The associated downstream task consists of classifying 23 cardiac abnormalities in a multi-label setting from reduced-lead ECGs. This dataset has been selected since it was used for similar evaluation in \textit{PhysioNet/Computing in Cardiology Challenge 2021} and in the reference work \cite{Oh2022_RLM}.
It consists of 9'458 12-lead ECGs, sampled at 500 Hz and lasting either 5 or 10 seconds. Each recording is associated with multiple (up to 7) cardiac abnormalities, for a total of 833 different label combinations. The dataset has been split on a subject basis into training, validation, and test sets according to an 80:10:10 ratio and label stratification. 
To evaluate model adaptability to various reduced-leads configurations, five different reduced-lead datasets have been created, by selecting the following leads from the original 12-lead dataset, as in \cite{Oh2022_RLM}: 12-lead, 6-lead (I, II, III, aVF, aVL, aVR), 3-lead (I, II, V2), 2-lead (I, II), and 1-lead (I).

\clearpage
\subsection{Experiments}
\label{sec:experiments}
\begin{figure}
  \centering
  \includegraphics[width=\textwidth]{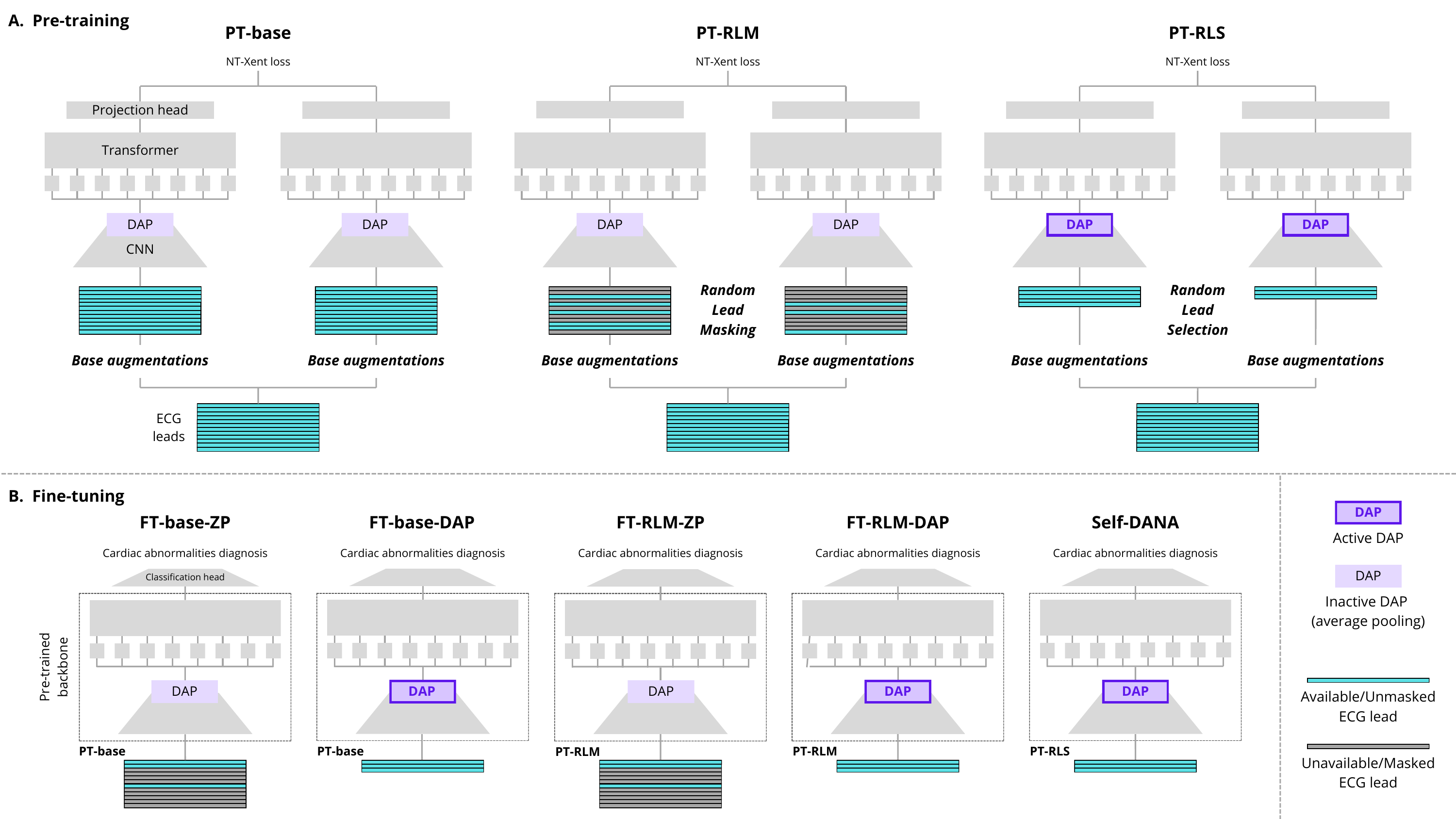}
  \caption{Overview of the experimental design, including the pre-training and fine-tuning procedures. The diagram illustrates how the Dimension Adaptive Pooling (DAP) layer and Random Lead Selection (RLS) components were integrated and evaluated across different experimental conditions.}
  \label{fig:experiments}
\end{figure}

Figure~\ref{fig:experiments} illustrates the experimental setup, detailing how the pre-training and fine-tuning procedures have been conducted. Below, we summarize three sets of experiments: \textit{(i)} assessing whether the DAP layer approach enhances resource efficiency while achieving performance comparable to the zero-padding technique; \textit{(ii)} evaluating the potential of RLS and the possible benefits of combining it with the DAP layer; \textit{(iii)} comparing the downstream performance of our channel-adaptive FM with the five different channel-specific supervised counterparts.

\paragraph{\textit{(i)} DAP layer}
We examined whether the adoption of the DAP layer, compared to the zero-padding technique, provided benefits in terms of resource efficiency without sacrificing performance. 
To this end, we pre-trained a baseline model (\textbf{PT-base}) using all 12 leads and creating the positive pairs exploiting only the \textit{base augmentations}. We then fine-tuned, on the five different reduced-leads datasets, two variants:
\begin{itemize}
    \item \textbf{FT-base-ZP}: zero-padding is applied to match the required input number of channels (12) for any reduced-lead configuration
    \item \textbf{FT-base-DAP}: the DAP layer is exploited to adapt the network to the available number of channels in the reduced-lead configuration
\end{itemize}

To ensure a fair comparison, both models were built using the same architecture. This means that the DAP layer is also present in \textbf{FT-base-ZP} but not active, i.e., it's used as a classic average pooling layer, since the input channels are always 12, as in the pre-training phase.

\paragraph{\textit{(ii)} Self-DANA}
We evaluated whether the combination with RLS augmentation, built ad-hoc for the DAP layer approach, was beneficial for it.
Additionally, we compared this setup with the RLM-based approach to ensure that the improved computational efficiency was not associated with a performance deterioration.
For this experiment, we pre-trained two other models:
\begin{itemize}
    \item \textbf{PT-RLM}: positive pairs for the contrastive learning task have been created by applying successively the \textit{base augmentations} and RLM.
    \item \textbf{PT-RLS}: positive pairs for the contrastive learning task have been created by applying successively the \textit{base augmentations} and RLS.
\end{itemize}
These models were then fine-tuned on the five different reduced-leads datasets using three different strategies:
\begin{itemize}
    \item \textbf{FT-RLM-ZP}: PT-RLM model has been fine-tuned on the downstream task, applying zero-padding to keep the number of channels always equal to 12, as in \cite{Oh2022_RLM}.
    \item \textbf{FT-RLM-DAP}: PT-RLM model has been fine-tuned on the downstream task without zero-padding, exploiting the DAP layer to keep the reduced number of channels.
    \item \textbf{Self-DANA}: PT-RLS model has been fine-tuned on the downstream task without zero-padding, exploiting the DAP layer to keep the reduced number of channels.
\end{itemize}
\textbf{FT-RLM-DAP} model has been evaluated to assess whether the RLS acts as an ad-hoc augmentation to enhance the DAP layer potential or whether another augmentation, such as RLM, would also lead to the same results.

\paragraph{(iii) Channel-adaptive FM vs channel-specific supervised models}
We investigate the benefit of fine-tuning a channel-adaptive FM on the desired task and lead configuration, rather than training different channel-specific models from scratch.
Specifically, for each of the five reduced-lead configurations, we trained a dedicated fully supervised model using only the available channels to solve the same cardiac abnormalities detection task on the Georgia dataset.

In all experiments described in points \textit{i} and \textit{ii}, full fine-tuning was carried out to address the cardiac abnormalities classification task. Details of the experimental settings are provided in the Appendix.

\paragraph{Evaluation}
The evaluation is focused on the fine-tuning phase, assessing both the computational efficiency and the performance of the fine-tuned models.\\ Performance on cardiac abnormalities diagnosis task is evaluated on the Georgia test set, through CinC score, the metric introduced ad-hoc by PhysioNet/Computing in Cardiology Challenge 2021 \cite{cinc21_1, cinc21_2}. It consists of a weighted accuracy that penalizes misdiagnosis with very different symptoms, reflecting the clinical reality that some misdiagnoses are more harmful than others. It ranges from 0 (an inactive classifier that always outputs the normal class) to 1 (a classifier that always outputs the true classes). For computation details, we refer to \cite{cinc21_1}.\\
Computational efficiency during the fine-tuning phase is evaluated in terms of peak memory usage and training time normalized by the number of epochs, for both CPU and GPU. The experiments have been performed on a cluster with an L40S GPU with 46068 MiB of RAM.\\
All the fine-tuning experiments have been repeated five times with different seeds to take into account results variability.

\section{Results and discussion}
\label{sec:results}
\subsection{Performance}

\paragraph{\textit{(i)} DAP layer} In Table~\ref{tab:exp_DAP_layer_perf} we evaluate whether the DAP layer approach (FT-base-DAP) would lead to performance not inferior to the zero-padding technique (FT-base-ZP). The average CinC score is clearly equal for the 12-lead configuration since the two techniques are equivalent by design in this setting. For all the other configurations, FT-base-DAP achieves comparable, if not slightly better, performance. These results demonstrate the feasibility of using the DAP layer as an alternative to zero-padding, since it does not lead to a performance drop.
\begin{table}[ht]
\caption{CinC score obtained in experiments \textit{(i)} on the five Georgia reduced-leads test sets. Results are reported as \textit{mean $\pm$ standard deviation}. Best results (highest CinC score) in bold.}
\label{tab:exp_DAP_layer_perf}
\centering
\begin{tabular*}{\textwidth}{@{\extracolsep\fill}lccccc}
\toprule
\multicolumn{1}{c}{} & \multicolumn{1}{c}{12 leads} & \multicolumn{1}{c}{6 leads} & \multicolumn{1}{c}{3 leads} & \multicolumn{1}{c}{2 leads} & \multicolumn{1}{c}{1 lead} \\
\midrule
FT-base-ZP & \textbf{\begin{tabular}[c]{@{}c@{}}\small0.600 $\pm$ 0.005\end{tabular}} & \begin{tabular}[c]{@{}c@{}}\small0.589 $\pm$ 0.004\end{tabular} & \begin{tabular}[c]{@{}c@{}}\small0.595 $\pm$ 0.009\end{tabular} & \begin{tabular}[c]{@{}c@{}}\small0.590 $\pm$ 0.006\end{tabular} & \begin{tabular}[c]{@{}c@{}}\small0.562 $\pm$ 0.004\end{tabular} \\
\textbf{\textbf{FT-base-DAP}} & \textbf{\begin{tabular}[c]{@{}c@{}}\small0.600 $\pm$ 0.005\end{tabular}} & \textbf{\begin{tabular}[c]{@{}c@{}}\small0.595 $\pm$ 0.003\end{tabular}} & \textbf{\begin{tabular}[c]{@{}c@{}}\small0.600 $\pm$ 0.005\end{tabular}} & \textbf{\begin{tabular}[c]{@{}c@{}}\small0.598 $\pm$ 0.005\end{tabular}} & \textbf{\begin{tabular}[c]{@{}c@{}}\small0.568 $\pm$ 0.005\end{tabular}} \\
\bottomrule
\end{tabular*}
\end{table}

\paragraph{\textit{(ii)} Self-DANA} In Table~\ref{tab:exp_selfDANA_perf} we evaluate the potential of RLS and the advantage of combining it with the DAP layer (Self-DANA). Comparing Self-DANA with FT-base-DAP, we evaluate the effect of adding RLS in the pre-training phase. The results show consistently higher performance with Self-DANA for all five configurations, demonstrating the beneficial effect of the proposed self-supervised dimension-adaptive pre-training. The performance also increases for the 12-lead configuration, indicating that the proposed RLS augmentation enhances model robustness even when using the pre-training channel configuration, highlighting its effectiveness as a general-purpose, independent ECG augmentation. The comparison with FT-RLM-DAP further confirms the customized nature of RLS for the DAP layer. Indeed, the combination of the DAP layer with RLM (FT-RLM-DAP) achieves lower performance than Self-DANA. Except for the 12-lead case, which is identical by design, FT-RLM-DAP also performs slightly worse than FT-RLM-ZP, especially for the single-lead setting. This suggests that the RLM technique needs to be combined with zero padding to better exploit its potential. Finally, the average CinC score of Self-DANA is comparable to, if not slightly higher than, FT-RLM-ZP. This further establishes our approach as a valid alternative to the RLM technique.

\begin{table}[ht]
\caption{CinC score obtained in experiments \textit{(ii)} on the five Georgia reduced-leads test sets. Results are reported as \textit{mean $\pm$ standard deviation}. Best results (highest CinC score) in bold.}
\label{tab:exp_selfDANA_perf}
\centering
\begin{tabular*}{\textwidth}{@{\extracolsep\fill}lccccc}
\toprule
\multicolumn{1}{c}{} & \multicolumn{1}{c}{12 leads} & \multicolumn{1}{c}{6 leads} & \multicolumn{1}{c}{3 leads} & \multicolumn{1}{c}{2 leads} & \multicolumn{1}{c}{1 lead} \\
\midrule
FT-base-DAP & \begin{tabular}[c]{@{}c@{}}\small0.600 $\pm$ 0.005\end{tabular} & \begin{tabular}[c]{@{}c@{}}\small0.595 $\pm$ 0.003\end{tabular} & \begin{tabular}[c]{@{}c@{}}\small0.600 $\pm$ 0.005\end{tabular} & \begin{tabular}[c]{@{}c@{}}\small0.598 $\pm$ 0.005\end{tabular} & \begin{tabular}[c]{@{}c@{}}\small0.568 $\pm$ 0.005\end{tabular} \\
\midrule
FT-RLM-ZP & \begin{tabular}[c]{@{}c@{}}\small0.612 $\pm$ 0.006\end{tabular} & \begin{tabular}[c]{@{}c@{}}\small0.606 $\pm$ 0.002\end{tabular} & \begin{tabular}[c]{@{}c@{}}\small0.610 $\pm$ 0.003\end{tabular} & \begin{tabular}[c]{@{}c@{}}\small0.611 $\pm$ 0.006\end{tabular} & \textbf{\begin{tabular}[c]{@{}c@{}}\small0.585 $\pm$ 0.004\end{tabular}} \\
FT-RLM-DAP & \begin{tabular}[c]{@{}c@{}}\small0.612 $\pm$ 0.006\end{tabular} & \begin{tabular}[c]{@{}c@{}}\small0.603 $\pm$ 0.007\end{tabular} & \begin{tabular}[c]{@{}c@{}}\small0.608 $\pm$ 0.003\end{tabular} & \begin{tabular}[c]{@{}c@{}}\small0.609 $\pm$ 0.005\end{tabular} & \begin{tabular}[c]{@{}c@{}}\small0.578 $\pm$ 0.003\end{tabular} \\
\midrule
\textbf{\textbf{Self-DANA}} & \textbf{\begin{tabular}[c]{@{}c@{}}\small0.619 $\pm$ 0.009\end{tabular}} & \textbf{\begin{tabular}[c]{@{}c@{}}\small0.613 $\pm$ 0.004\end{tabular}} & \textbf{\begin{tabular}[c]{@{}c@{}}\small0.617 $\pm$ 0.009\end{tabular}} & \textbf{\begin{tabular}[c]{@{}c@{}}\small0.618 $\pm$ 0.008\end{tabular}} & \begin{tabular}[c]{@{}c@{}}\small0.583 $\pm$ 0.004\end{tabular} \\
\bottomrule
\end{tabular*}
\end{table}
To provide context for the CinC scores obtained in our experiments, we report reference values from the literature for this type of task and dataset. However, we underline that directly comparing our performance with the literature could not provide a fair assessment, as test datasets and conditions were slightly different. Moreover, contrary to our case, both the models from the literature have been pre-trained also on the Georgia dataset, making them in-domain evaluations.
The performance obtained by the reference work \cite{Oh2022_RLM} using SimCLR framework with RLM on Georgia and CPSC test sets combined for the same task are the following (reported as mean $\pm$ 95\% confidence interval across three seeds): $0.578 \pm 0.015$ for 12-lead, $0.497 \pm 0.002$ for 6-lead, $0.535 \pm 0.015$ for 3-lead, $0.484 \pm 0.004$ for 2-lead and $0.393 \pm 0.012$ for 1-lead configurations. Additionally, the CinC score obtained by the winning team \cite{Nejedly2021_cinc21_winner} of the \textit{PhysioNet/Computing in Cardiology Challenge 2021} on the (unavailable) Georgia test set is 0.61. This value refers to the \textit{all-lead} configuration, computed as the mean between 12-lead, 3-lead, and 2-lead scores.

\paragraph{(iii) Channel-adaptive FM vs channel-specific supervised models}
In Table~\ref{tab:exp_sup}, we compare our channel-adaptive approach with the 5 channel-specific supervised models, each trained directly with the target reduced-lead configuration. The results show that Self-DANA consistently outperforms the supervised counterpart across all five configurations. This further supports that, to tackle tasks with reduced-lead configuration, it is beneficial to exploit a channel-adaptive FM, robust to different lead configurations, rather than the channel-specific supervised counterpart.

\begin{table}[ht]
\caption{CinC score obtained in experiments \textit{(iii)} on the five Georgia reduced-leads test sets. Results are reported as \textit{mean $\pm$ standard deviation}. Best results (highest CinC score) in bold.}
\label{tab:exp_sup}
\centering
\begin{tabular*}{\textwidth}{@{\extracolsep\fill}lccccc}
\toprule
\multicolumn{1}{c}{} & \multicolumn{1}{c}{12 leads} & \multicolumn{1}{c}{6 leads} & \multicolumn{1}{c}{3 leads} & \multicolumn{1}{c}{2 leads} & \multicolumn{1}{c}{1 lead} \\
\midrule
Supervised & \small0.578 ± 0.006 & \small0.573 ± 0.007 & \small0.574 ± 0.005 & \small0.581 ± 0.002 & \small0.547 ± 0.003 \\
\textbf{\textbf{Self-DANA}} & \textbf{\begin{tabular}[c]{@{}c@{}}\small0.619 $\pm$ 0.009\end{tabular}} & \textbf{\begin{tabular}[c]{@{}c@{}}\small0.613 $\pm$ 0.004\end{tabular}} & \textbf{\begin{tabular}[c]{@{}c@{}}\small0.617 $\pm$ 0.009\end{tabular}} & \textbf{\begin{tabular}[c]{@{}c@{}}\small0.618 $\pm$ 0.008\end{tabular}} & \textbf{\begin{tabular}[c]{@{}c@{}}\small0.583 $\pm$ 0.004\end{tabular}} \\
\bottomrule
\end{tabular*}
\end{table}

\subsection{Resource efficiency}
\label{sec:results_eff}
We compared Self-DANA and the reference FT-RLM-ZP in terms of resource efficiency, i.e., memory and time consumption, reported in Table~\ref{tab:memory_eff} and \ref{tab:time_eff}, respectively. The peak GPU memory required by Self-DANA is considerably lower for all the reduced-lead configurations and, as expected, decreases with the number of available channels. Hence, the percentage of memory that can be saved with our approach ranges from 18.76\% with 6 leads to 34.41\% with a single lead. Similarly, a substantial reduction in CPU memory usage has been observed, amounting to 49.91\% or 69.32\%, depending on the channels configuration.
These results highlight the superior efficiency of our approach in terms of memory consumption. This is directly attributable to the elimination of zero-padding, made possible by the introduction of the DAP layer.\\
In Table~\ref{tab:time_eff}, we show that the average time needed per training epoch is lower with our approach in all configurations except the 12-lead case, for which it is comparable.
Interestingly, the standard deviations of both CPU and GPU times across the five runs are considerably lower with Self-DANA, indicating greater stability in the results.

\begin{table}[ht]
\caption{Peak CPU and GPU memory required for FT-RLM-ZP and Self-DANA during fine-tuning. The percentage difference between Self-DANA and FT-RLM-ZP memory consumption (memory saving percentage) is also reported. Best results (lowest peak memory) in bold.}
\label{tab:memory_eff}
\centering
\begin{tabular*}{\textwidth}{@{\extracolsep\fill}lccccc}
\toprule
\multicolumn{1}{c}{} & \multicolumn{1}{c}{12 leads} & \multicolumn{1}{c}{6 leads} & \multicolumn{1}{c}{3 leads} & \multicolumn{1}{c}{2 leads} & \multicolumn{1}{c}{1 lead} \\
\midrule
\multicolumn{6}{c}{Peak CPU memory (MB)}\\
\midrule
FT-RLM-ZP & \textbf{\begin{tabular}[c]{@{}c@{}}\small58.68\end{tabular}} & \begin{tabular}[c]{@{}c@{}}\small58.68\end{tabular} & \begin{tabular}[c]{@{}c@{}}\small58.68\end{tabular} & \begin{tabular}[c]{@{}c@{}}\small58.68\end{tabular} & \begin{tabular}[c]{@{}c@{}}\small58.68\end{tabular} \\
\textbf{\textbf{Self-DANA}} & \textbf{\begin{tabular}[c]{@{}c@{}}\small58.68\end{tabular}} & \textbf{\begin{tabular}[c]{@{}c@{}}\small29.39\end{tabular}} & \textbf{\begin{tabular}[c]{@{}c@{}}\small18.00\end{tabular}} & \textbf{\begin{tabular}[c]{@{}c@{}}\small18.00\end{tabular}} & \textbf{\begin{tabular}[c]{@{}c@{}}\small18.00\end{tabular}} \\
\midrule
Memory saving & \textbf{\begin{tabular}[c]{@{}c@{}}\small-0.00\%\end{tabular}} & \textbf{\begin{tabular}[c]{@{}c@{}}\small-49.91\%\end{tabular}} & \textbf{\begin{tabular}[c]{@{}c@{}}\small-69.32\%\end{tabular}} & \textbf{\begin{tabular}[c]{@{}c@{}}\small-69.32\%\end{tabular}} & \textbf{\begin{tabular}[c]{@{}c@{}}\small-69.32\%\end{tabular}} \\
\midrule
\multicolumn{6}{c}{Peak GPU memory (GB)}\\
\midrule
FT-RLM-ZP & \textbf{\begin{tabular}[c]{@{}c@{}}\small23.19\end{tabular}} & \begin{tabular}[c]{@{}c@{}}\small23.19\end{tabular} & \begin{tabular}[c]{@{}c@{}}\small23.19\end{tabular} & \begin{tabular}[c]{@{}c@{}}\small23.19\end{tabular} & \begin{tabular}[c]{@{}c@{}}\small23.19\end{tabular} \\
\textbf{\textbf{Self-DANA}} & \textbf{\begin{tabular}[c]{@{}c@{}}\small23.19\end{tabular}} & \textbf{\begin{tabular}[c]{@{}c@{}}\small18.84\end{tabular}} & \textbf{\begin{tabular}[c]{@{}c@{}}\small16.66\end{tabular}} & \textbf{\begin{tabular}[c]{@{}c@{}}\small15.93\end{tabular}} & \textbf{\begin{tabular}[c]{@{}c@{}}\small15.21\end{tabular}} \\
\midrule
Memory saving & \textbf{\begin{tabular}[c]{@{}c@{}}\small-0.00\%\end{tabular}} & \textbf{\begin{tabular}[c]{@{}c@{}}\small-18.76\%\end{tabular}} & \textbf{\begin{tabular}[c]{@{}c@{}}\small-28.16\%\end{tabular}} & \textbf{\begin{tabular}[c]{@{}c@{}}\small-31.31\%\end{tabular}} & \textbf{\begin{tabular}[c]{@{}c@{}}\small-34.41\%\end{tabular}} \\
\bottomrule
\end{tabular*}
\end{table}

\begin{table}[ht]
\caption{Average epoch training time measured for FT-RLM-ZP and Self-DANA during fine-tuning. Results are reported as \textit{mean $\pm$ standard deviation}. Best results (lowest normalized time) in bold.}
\label{tab:time_eff}
\centering
\begin{tabular*}{\textwidth}{@{\extracolsep\fill}lccccc}
\toprule
\multicolumn{1}{c}{} & \multicolumn{1}{c}{12 leads} & \multicolumn{1}{c}{6 leads} & \multicolumn{1}{c}{3 leads} & \multicolumn{1}{c}{2 leads} & \multicolumn{1}{c}{1 lead} \\
\midrule
\multicolumn{6}{c}{Average epoch CPU time (s)}\\
\midrule
FT-RLM-ZP & \textbf{\begin{tabular}[c]{@{}c@{}}\small137.71 $\pm$ 6.59\end{tabular}} & \begin{tabular}[c]{@{}c@{}}\small123.50 $\pm$ 11.28\end{tabular} & \begin{tabular}[c]{@{}c@{}}\small123.56 $\pm$ 10.52\end{tabular} & \begin{tabular}[c]{@{}c@{}}\small119.39 $\pm$ 7.34\end{tabular} & \begin{tabular}[c]{@{}c@{}}\small118.82 $\pm$ 8.92\end{tabular} \\
\textbf{\textbf{Self-DANA}} & \begin{tabular}[c]{@{}c@{}}\small140.43 $\pm$ 3.74\end{tabular} & \textbf{\begin{tabular}[c]{@{}c@{}}\small116.37 $\pm$ 0.89\end{tabular}} & \textbf{\begin{tabular}[c]{@{}c@{}}\small105.95 $\pm$ 0.76\end{tabular}} & \textbf{\begin{tabular}[c]{@{}c@{}}\small102.72 $\pm$ 0.79\end{tabular}} & \textbf{\begin{tabular}[c]{@{}c@{}}\small98.84 $\pm$ 0.88\end{tabular}} \\
\midrule
\multicolumn{6}{c}{Average epoch GPU time (s)}\\
\midrule
FT-RLM-ZP & \textbf{\begin{tabular}[c]{@{}c@{}}\small139.58 $\pm$ 0.97\end{tabular}} & \begin{tabular}[c]{@{}c@{}}\small139.81 $\pm$ 0.54\end{tabular} & \begin{tabular}[c]{@{}c@{}}\small139.88 $\pm$ 0.37\end{tabular} & \begin{tabular}[c]{@{}c@{}}\small139.68 $\pm$ 0.49\end{tabular} & \begin{tabular}[c]{@{}c@{}}\small140.00 $\pm$ 0.41\end{tabular} \\
\textbf{\textbf{Self-DANA}} & \begin{tabular}[c]{@{}c@{}}\small140.13 $\pm$ 0.10\end{tabular} & \textbf{\begin{tabular}[c]{@{}c@{}}\small122.08 $\pm$ 0.08\end{tabular}} & \textbf{\begin{tabular}[c]{@{}c@{}}\small113.10 $\pm$ 0.16\end{tabular}} & \textbf{\begin{tabular}[c]{@{}c@{}}\small110.17 $\pm$ 0.22\end{tabular}} & \textbf{\begin{tabular}[c]{@{}c@{}}\small107.43 $\pm$ 0.12\end{tabular}} \\
\bottomrule
\end{tabular*}
\end{table}

\section{Conclusions}
\label{sec:conclusions}
We introduce Self-DANA, a novel, easy-to-integrate solution for ECG FMs that makes them adaptable to reduced-channel scenarios. Our approach achieves performance comparable to the state-of-the-art while considerably increasing the efficiency in terms of training time and memory resources required. This is possible thanks to the combination of the DAP layer, which serves as an efficient alternative to zero-padding, and RLS, our new augmentation tailored to enhance channel adaptability during pre-training. Our approach is also easy to implement and seamlessly integrates into existing CL-based FMs to improve their generalizability to an extended range of reduced-channel downstream scenarios. The dual benefits of channel adaptability and resource efficiency position Self-DANA as a compelling option for real-world applications, especially in the portable and wearable technology field.\\
Despite these promising outcomes, this study has certain limitations. First, Self-DANA has been validated only for SSCL frameworks, which, however, represent the great majority of the ones proposed for ECG signals. Moreover, we believe that our approach could also be easily adapted to non-CL frameworks, e.g., by using RLS as a general-purpose augmentation during pre-training to make the models more robust to reduced-channel scenarios. This could be investigated and evaluated as future work.
Second, the proposed approach has been tested only on a single downstream dataset and task, even if the Georgia dataset covers a wide range of cardiac abnormalities, and it has been used in the literature for similar evaluations. We emphasize that this study serves as a proof of concept for our approach, and future exploration on different architectures, tasks, and datasets would be of significant interest.
Third, we did not compare Self-DANA performance with the five channel-specific self-supervised counterparts, i.e., a different FM for each of the five reduced-lead configurations, as it conflicts with the principle of generalizability, which is central to FMs, while building a single adaptable FM is ultimately more efficient.\\
To conclude, given that reduced-channel configurations are a common challenge across many biosignals and wearable technologies, we believe Self-DANA has the potential for broader impact. We plan to extend our evaluation to other signal modalities and multi-modal applications, further exploring its scalability and adaptability in real-world healthcare applications.

\clearpage







\newcolumntype{C}[1]{>{\centering\arraybackslash}p{#1}}

\renewcommand{\thefigure}{A\arabic{figure}}
\renewcommand{\thetable}{A\arabic{table}}
\setcounter{figure}{0}
\setcounter{table}{0}

\appendix

\section{Technical Appendices and Supplementary Material}
\subsection{Datasets}
\label{app:datasets}
This section supplies further information about the datasets exploited in pre-training and fine-tuning experiments. Then, the datasets' analyses in terms of distributions of recordings and labels across the split sets, the preprocessing applied, and the exact splits subdivision are also provided.

\paragraph{CPSC and CPSC-extra}
\textit{CPSC and CPSC-extra} datasets, from \textit{PhysioNet/Computing in Cardiology Challenge 2021}, derive from the China Physiological Signal Challenge in 2018 (CPSC2018). It consists of two sets of 6,877 (male: 3,699; female: 3,178) and 3,453 (male: 1,843; female: 1,610) of 12-ECG recordings lasting from 6 seconds to 60 seconds. Each recording is sampled at 500 Hz.

\paragraph{INCART} \textit{St Petersburg INCART 12-lead Arrhythmia Database (INCART)} dataset, from \textit{PhysioNet/Computing in Cardiology Challenge 2021}, consists of 74 annotated recordings extracted from 32 Holter records. Each record is 30 minutes long and contains 12 standard leads, each sampled at 257 Hz.

\paragraph{Chapman-Shaoxing and Ningbo} \textit{Chapman University, Shaoxing People’s Hospital (Chapman-Shaoxing)} dataset and \textit{Ningbo First Hospital (Ningbo)} dataset, from \textit{PhysioNet/Computing in Cardiology Challenge 2021}, contain a total of 45,152 ECGs (all shared as training data). Each recording is 10 second long with a sampling frequency of 500 Hz.

\paragraph{Code-15\%}
\textit{CODE-15\%} dataset is a subset of the larger CODE dataset, created through stratified sampling to include 15\% of the patients. It consists of 345,779 annotated 12-lead ECG exams from 233,770 patients. The data was collected by the Telehealth Network of Minas Gerais (TNMG), a public telehealth system serving most municipalities in Minas Gerais, Brazil, between 2010 and 2016.

\paragraph{Off-test}
\textit{Offline Test Set of ECG Multi-label Classification (Off-test} dataset is the offline test set for the study "Practical arrhythmias detection algorithm for wearable 12-lead ECG via self-supervised learning on large-scale dataset." It includes 7,000 12-lead wearable ECG recordings, each 15 seconds long with a sampling frequency of 500 Hz. The data covers 60 rhythm classes, all reviewed and diagnosed by cardiologists and spanning a wide range of normal and abnormal heart conditions.

\paragraph{Georgia}
\textit{Georgia} dataset, from \textit{PhysioNet/Computing in Cardiology Challenge 2021}, represents a unique demographic of the Southeastern United States. It includes 20,672 ECG recordings, with 10,344 used for training, 5,167 for validation, and 5,161 for testing. Each ECG lasts between 5 and 10 seconds and is sampled at a frequency of 500 Hz.\\
We used just the training set of this dataset since it is the only one made available. It has been split on a subject basis into training, validation, and test sets according to an 80:10:10 ratio and label stratification.

\paragraph{Preprocessing}
All pre-training and fine-tuning datasets have been prepared using the same procedure: resampling at 500 Hz, segmentation into non-overlapping 5s windows, and pre-processing.
First, resampling was done, when needed, to a frequency of 500 Hz. Then, each recording has been segmented into non-overlapping 5s-windows. Finally, each window has been pre-processed by removing the mean and applying a 5th-order moving average filter and a Butterworth 4th-order band-pass filter with cutoff frequencies of 0.5 and 40 Hz.

\paragraph{Splits}
All the splits have been performed on a subject basis to ensure all the recordings of a subject fall in the same split set. 

\subsection{Base augmentations}
\label{app:augmentations}
This section provides a description and parameters related to the five augmentations exploited in the \textit{base augmentations} sequence.

\paragraph{Amplitude scaling} It multiplies the signal's amplitude by a random scaling factor $s$. This trains the model to be invariant to amplitude differences that may arise from patient-specific variations or electrode placement, encouraging the encoder to focus on morphological patterns and temporal structure rather than absolute signal magnitude.
Based on the results of \cite{Soltanieh2022_aug}, we randomly select, for each ECG window, a scaling factor $s$ in the range of 0.5 to 1.7, meaning that the ECG will be rescaled between 50\% and 170\% of its original amplitude.

\paragraph{Gaussian noise} It simulates the noise inherently present in real-world settings, due to electrode movement, muscle artifacts, etc. The augmented ECG view is obtained by adding Gaussian noise  $\epsilon \sim \mathcal{N}(0, \sigma^2)$.
Based on the results of \cite{Soltanieh2022_aug}, we randomly select, for each ECG window, a standard deviation value $\sigma$ in the range of 0.1 to 0.25, introducing minor variability while preserving the core structure of the signal.

\paragraph{Crop and resize} 
It consists of cropping a random contiguous portion of the ECG signal and then resizing it back to the original length via interpolation. This transformation changes the temporal resolution but retains overall shape information.
Following \cite{Mehari2022_SSLECG}, a portion of the signal, with a randomly determined length between 50\% and 100\% of the original signal, is randomly cropped. The cropped segment is then resized to the target length using cubic spline interpolation.

\paragraph{Time masking}
It emulates signal dropout or corruption by setting a random portion of the signal to zero. This improves the model's robustness to missing or noisy data, forcing it to reconstruct or interpret partial signals.
Based on the results of \cite{Soltanieh2022_aug, Mehari2022_SSLECG}, we randomly select the masking ratio within the range of 0\% to 50\% of the signal length and then mask with zeros a random portion of the desired length.

\paragraph{Time warping}
It transforms the temporal structure of signals by stretching or squeezing random segments, helping the model handle variations in heart rate and temporal dynamics.
A random number of segments, ranging from 4 to 9, is first selected. These segments are then randomly designated to be either stretched by a factor of 2 or squeezed by a factor of 0.5. To match the original length after these temporal modifications, resampling is performed using piecewise cubic Hermite interpolating polynomials (PCHIP). For the implementation we refer to \cite{selfeeg_code, selfeeg_paper}. 

\clearpage

\subsection{Architectural details}
\label{app:architecture}

\begin{figure}[ht]
  \centering
  \includegraphics[width=0.4\textwidth]{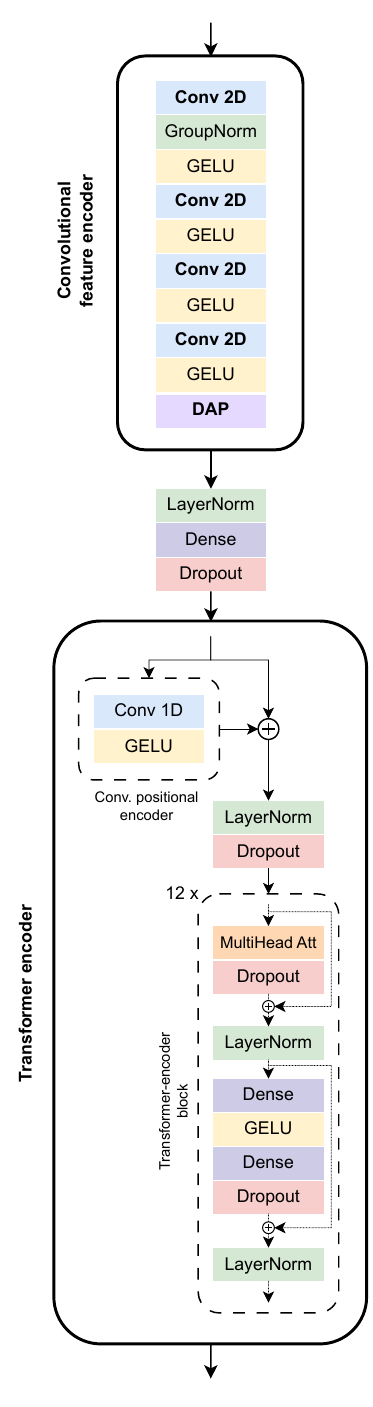}
  \caption{Graphical representation of backbone architecture.}
  \label{app_fig:architecture}
\end{figure}

\paragraph{Backbone}
A detailed scheme of the backbone architecture with hyperparameters is provided in Table~\ref{app_tab:architecture}, while a graphical overview is shown in Figure~\ref{app_fig:architecture}.

\begin{table}[ht]
\caption{Detailed backbone architecture with hyperparameters.}
\label{app_tab:architecture}
\begin{tabular}{p{3cm}c C{2.5cm}}
\toprule
\textbf{Layer}           & \textbf{Hyperparameters}                            & \textbf{Output Shape} \\
\midrule
\textit{input}           &     -                                                & {[}B, C, 2500{]}      \\
\midrule
& \textit{\textbf{Convolutional feature encoder}} & \\
\midrule
Conv2D                   & k=(1,2), s=(1,2)                                    & {[}B, 256, C, 1250{]} \\
GroupNorm                & n\_groups=256, eps=1e-5                             & {[}B, 256, C, 1250{]} \\
GELU                     &  -                                                 & {[}B, 256, C, 1250{]} \\
Conv2D                   & k=(1,2), s=(1,2)                                    & {[}B, 256, C, 625{]}  \\
GELU                     &    -                                               & {[}B, 256, C, 625{]}  \\
Conv2D                   & k=(1,2), s=(1,2)                                    & {[}B, 256, C, 312{]}  \\
GELU                     &  -                                                 & {[}B, 256, C, 312{]}  \\
Conv2D                   & k=(1,2), s=(1,2)                                    & {[}B, 256, C, 156{]}  \\
GELU                     &   -                                                & {[}B, 256, C, 156{]}  \\
Adaptive Avg Pool & out\_dim=(1,156)                                    & {[}B, 256, 1, 156{]}  \\
Flatten                  & dim=2                                               & {[}B, 256, 156{]}     \\
\midrule
Transpose                &   -                                                  & {[}B, 156, 256{]}     \\
LayerNorm                & norm\_shape=256, eps=1e-05                          & {[}B, 156, 256{]}     \\
Linear                   & in\_dim=256, out\_dim=768                           & {[}B, 156, 768{]}     \\
Dropout                  & p=0.1                                               & {[}B, 156, 768{]}     \\
\midrule
& \textit{\textbf{Convolutional positional encoder}} & \\
\midrule
Conv1D                   & k=128, s=1, pad=64, groups=16                       & {[}B, 768, 157{]}     \\
Padding                  &     -                                                & {[}B, 768, 156{]}     \\
GELU                     &    -                                               & {[}B, 768, 156{]}     \\
Transpose                &     -                                                & {[}B, 156, 768{]}     \\
\midrule
& \textit{\textbf{Transformer encoder}} & \\
\midrule
LayerNorm                & norm\_shape=768, eps=1e-05                          & {[}B, 156, 768{]}     \\
Dropout                  & p=0.1                                               & {[}B, 156, 768{]}     \\
Transpose                &  -                                                   & {[}156, B, 768{]}     \\
\midrule
& \textit{\textbf{Transformer-encoder blocks (x 12)}} & \\
\midrule
MultiHead Attention      & embed\_dim=768, n\_heads=12, dropout=0.1, q=k=v=768 & {[}156, B, 768{]}     \\
Dropout                  & p=0.1                                               & {[}156, B, 768{]}     \\
LayerNorm                & norm\_shape=768, eps=1e-05                          & {[}156, B, 768{]}     \\
Linear                   & in\_dim=768, out\_dim=3072                          & {[}156, B, 3072{]}    \\
GELU                     &   -                                                & {[}156, B, 3072{]}    \\
Linear                   & in\_dim=3072, out\_dim=768                          & {[}156, B, 768{]}     \\
Dropout                  & p=0.1                                               & {[}156, B, 768{]}     \\
LayerNorm                & norm\_shape=768, eps=1e-05                          & {[}156, B, 768{]}     \\
Transpose                &    -                                                 & {[}B, 156, 768{]}    \\
\bottomrule
\end{tabular}
\end{table}

\clearpage

\paragraph{Pre-training}
For the pre-training phase, we exploited a projection head consisting of a fully connected layer with input dimension 768 and output dimension 256, followed by a 1D batch normalization layer applied to the 256 output neurons, as in \cite{Oh2022_RLM}. The backbone comprises 90'367'616 parameters, and the projection head contributes an additional 197'376, resulting in a total parameter count of 90'564'992.\\
For the self-supervised pre-training with the SimCLR framework, we employed the original loss function, i.e., the \textit{NT-Xent loss} function, defined as
\begin{center}
    \begin{equation}
        \ell_{i,j} = -\log \frac{\exp\left(\frac{\mathrm{sim}(\mathbf{z}_i, \mathbf{z}_j)}{\tau}\right)}{\sum\limits_{k=1}^{2N} \mathbf{1}_{[k \ne i]} \exp\left(\frac{\mathrm{sim}(\mathbf{z}_i, \mathbf{z}_k)}{\tau}\right)}
    \end{equation}
\end{center}
where \(\mathbf{z}_i\) and \(\mathbf{z}_j\) are the projected representations of the positive pair, \(\mathrm{sim}(\mathbf{u}, \mathbf{v}) = \frac{\mathbf{u}^\top \mathbf{v}}{\|\mathbf{u}\| \|\mathbf{v}\|}\) is the cosine similarity, \(\tau\) is a temperature parameter, and \(N\) is the batch size.\\
Specifically, we set \(\tau=0.5\) and \(N=128\).

\paragraph{Fine-tuning}
For the fine-tuning phase, we replaced the projection head with a classification head composed of a fully connected layer with input dimension 768 and output dimension 23 (number of classes of the downstream task) and a sigmoid activation function. The classification head accounts for a total of 796 parameters.\\
To address the cardiac abnormality classification task, where multiple labels can be associated with a single recording, we employed a binary cross-entropy loss function applied independently to each of the 23 output neurons, enabling the model to learn the presence or absence of each class.

\paragraph{Source code}
For the implementation of the encoder and projection head architecture, we refer to the source code provided in \cite{fairseq-signals_code}, reporting the architecture exploited in \cite{Oh2022_RLM} and \cite{McKeen2024_ECGFM}. Unless otherwise specified in this Appendix, we kept the same hyperparameters as in the original work and code.
The rest of the code structure, including the SimCLR framework, builds upon the selfEEG library \cite{selfeeg_code, selfeeg_paper}.

\subsection{Experimental settings}
\label{app:train_detail}
\paragraph{Training details}
During the pre-training phase, we employed the Adam optimizer with an exponential decay setting an initial learning rate $lr = 5e-5$ and a decay factor $\gamma=0.97$. We trained our models for a maximum of 100 epochs, applying early stopping based on validation loss with a patience of $10$ and a minimum improvement threshold of $1e-5$.
During the fine-tuning phase, we employed the Adam optimizer with an exponential decay setting an initial learning rate $lr = 1e-5$ and a decay factor $\gamma=0.97$. We trained our models for a maximum of 50 epochs, applying early stopping based on validation loss with a patience of $10$ and a minimum improvement threshold of $1e-3$. We repeated the fine-tuning five times with different seeds $(0, 1, 2, 3, 4)$.
\paragraph{Data sampling strategies}
For batch generation, we exploited two different sampling strategies for the pre-training and fine-tuning phases.
During pre-training, we employed a batch size of $128$ 5s-segments. We populate the batches by sampling the segments from the whole pre-training set with uniform distribution.
During fine-tuning, we employed a batch size of $128$ 5s-segments. To balance the labels distribution in the batch, we first select one of the 23 cardiac abnormalities with random uniform sampling (\textit{class sampling}); we then randomly chose a subject (\textit{subject sampling}) and a recording (\textit{recording sampling}) containing the desired label; and, finally, we randomly select a 5s-segment from that recording. An epoch is concluded after $Nb=Tr/N$ batches, where $Tr$ is equal to the number of 5s segments in the train set, and $N$ is the batch size.

\clearpage
\subsection{Supplementary results}
\label{app:results}

\begin{figure}[ht]
  \centering
  \includegraphics[width=\textwidth]{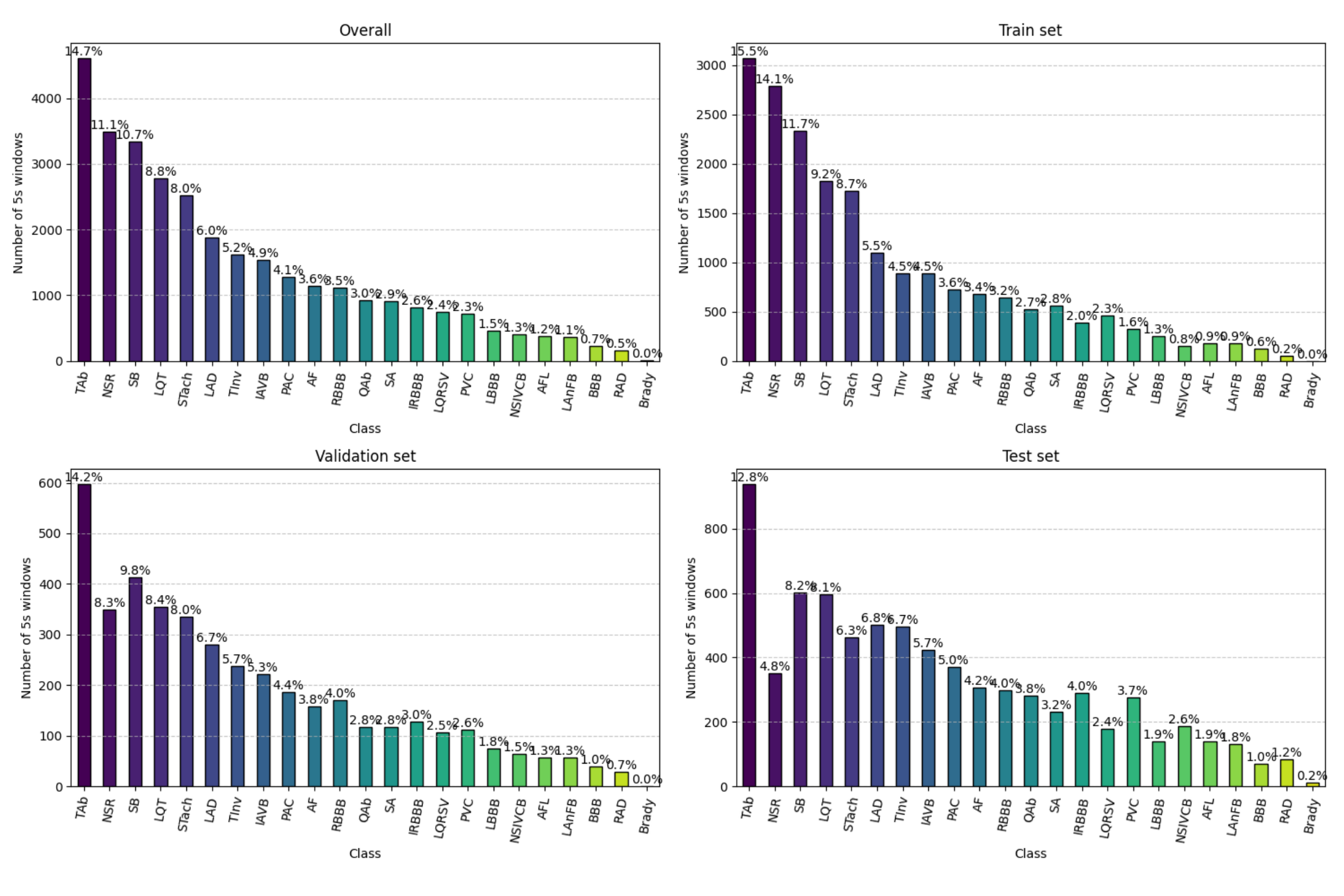}
  \caption{Label distribution of the Georgia dataset and in its train, validation, and test subsets. Labels' acronyms are described in Table \ref{app_tab:georgia_labels}.}
  \label{app_fig:georgia_db}
\end{figure}

\paragraph{Datasets analysis}
Table~\ref{app_tab:pretrain_db} and \ref{app_tab:finetune_db} report the number of subjects, recordings, and 5s windows for each split set of pre-training and fine-tuning datasets, respectively.
The label distribution in the Georgia dataset and its corresponding splits is presented in two forms: by counting the occurrence of each of the 23 individual labels across recordings (see Table~\ref{app_tab:georgia_labels} and Figure~\ref{app_fig:georgia_db}), and by analyzing the 833 unique label combinations assigned to the recordings (see Table~\ref{app_tab:georgia_labels_comb}).

\begin{table}[ht]
\caption{Number of subjects (\textit{sbj}), recordings (\textit{rec}), and 5s windows (\textit{win}) in training and validation sets of the pre-training datasets}
\label{app_tab:pretrain_db}
\centering
\begin{tabular*}{\textwidth}{@{\extracolsep\fill}lcccccc}
\toprule
                   & \multicolumn{3}{c}{Train}                      & \multicolumn{3}{c}{Validation}                 \\
\midrule
\textit{}          & \textit{sbj} & \textit{rec} & \textit{win} & \textit{sbj} & \textit{rec} & \textit{win} \\
\midrule
Code-15\% & 186131        & 274850          & 549700       & 46601         & 68712           & 137424       \\
Ningbo             & 27923         & 27923           & 55846        & 6981          & 6981            & 13962  \\ 
INCART             & 25            & 59              & 21240        & 7             & 15              & 5400         \\
Off-test  & 5600          & 5600            & 16800        & 1400          & 1400            & 4200         \\
Chapman-Shaoxing            & 8198          & 8198            & 16396        & 2049          & 2049            & 4098         \\
CPSC               & 5493          & 5493            & 16138        & 1384          & 1384            & 4024         \\
CPSC-extra         & 2741          & 2741            & 8175         & 712           & 712             & 2021         \\
\midrule
Total              & 236111 & 324864 & 684295 & 59134 & 81253 & 171129\\
\bottomrule
\end{tabular*}
\end{table}

\begin{table}[ht]
\caption{Number of subjects (\textit{sbj}), recordings (\textit{rec}), and 5s windows (\textit{win}) in training, validation and test sets of the fine-tuning dataset}
\label{app_tab:finetune_db}
\centering
\begin{tabular*}{\textwidth}{@{\extracolsep\fill}lccccccccc}
\toprule
\multicolumn{1}{l}{} & \multicolumn{3}{c}{Train} & \multicolumn{3}{c}{Validation} & \multicolumn{3}{c}{Test} \\
\midrule
            & \textit{sbj} & \textit{rec} & \textit{win} & \textit{sbj} & \textit{rec} & \textit{win} & \textit{sbj} & \textit{rec} & \textit{win} \\
\midrule
Georgia     & 6745   & 6745   & 13450   & 1147     & 1147     & 2292     & 1566   & 1566   & 3130   \\
\bottomrule
\end{tabular*}
\end{table}

\begin{table}[ht]
\caption{Number of 5s windows in which each label is present in train, validation, test, and whole Georgia dataset. A description of the label acronym is also provided.}
\label{app_tab:georgia_labels}
\centering
\begin{tabular}{llcccc}
\toprule
\textbf{Label} & \textbf{Description} & \textbf{All} & \textbf{Train} & \textbf{Validation} & \textbf{Test} \\
\midrule
TAb      & \small T wave abnormal                                  & 4605 & 3074 & 598 & 933 \\
NSR      & \small Normal sinus rhythm                              & 3487 & 2787 & 350 & 350 \\
SB       & \small Sinus bradycardia                                & 3345 & 2331 & 414 & 600 \\
LQT      & \small Prolonged QT interval                            & 2774 & 1825 & 355 & 594 \\
STach    & \small Sinus tachycardia                                & 2522 & 1728 & 336 & 458 \\
LAD      & \small Left axis deviation                              & 1878 & 1098 & 280 & 500 \\
TInv     & \small T wave inversion                                 & 1620 & 889  & 238 & 493 \\
IAVB     & \small First degree av block                            & 1535 & 891  & 224 & 420 \\
PAC      & \small Premature atrial contraction                     & 1277 & 720  & 186 & 371 \\
AF       & \small Atrial fibrillation                              & 1138 & 676  & 158 & 304 \\
RBBB     & \small Right bundle branch block                        & 1110 & 647  & 170 & 293 \\
QAb      & \small Q wave abnormal                                  & 927  & 527  & 118 & 282 \\
SA       & \small Sinus arrhythmia                                 & 909  & 559  & 118 & 232 \\
IRBBB    & \small Incomplete right bundle branch block             & 809  & 392  & 128 & 289 \\
LQRSV    & \small Low QRS voltages                &  747  & 463  & 108 & 176 \\
PVC      & \small Premature ventricular contractions               & 713  & 326  & 111 & 276 \\
LBBB     & \small Left bundle branch block                         & 462  & 250  & 74  & 138 \\
NSIVCB   & \small Nonspecific intraventricular conduction disorder & 406  & 154  & 64  & 188 \\
AFL      & \small Atrial flutter                                   & 372  & 178  & 56  & 138 \\
LAnFB    & \small Left anterior fascicular block                   & 360  & 174  & 56  & 130 \\
BBB      & \small Bundle branch block                              & 231  & 120  & 40  & 71  \\
RAD      & \small Right axis deviation                             & 163  & 49   & 29  & 85  \\
Brady    & \small Bradycardia                                      & 12   & 6    & 2   & 4   \\
\bottomrule
\end{tabular}
\end{table}

\begin{table}[ht]
\caption{Number of 5s windows for each set of labels in training, validation, test, and whole Georgia dataset. Only the first and the last 25 more frequent label combinations (out of 833) are reported here.} 
\label{app_tab:georgia_labels_comb}
\centering
\begin{tabular}{lcccc}
\toprule
\textbf{Set of labels} & \textbf{All} & \textbf{Train} & \textbf{Validation} & \textbf{Test} \\
\midrule
NSR              & 3487          & 2787         & 350            & 350      \\
SB               & 1433          & 1145         & 144            & 144      \\
STach            & 1016          & 812          & 102            & 102      \\
TAb              & 893           & 713          & 90             & 90       \\
LQT              & 808           & 644          & 82             & 82       \\
SA               & 386           & 306          & 40             & 40       \\
LQT, TAb         & 384           & 304          & 40             & 40       \\
LAD              & 340           & 272          & 34             & 34       \\
PAC              & 322           & 254          & 34             & 34       \\
TInv             & 288           & 228          & 30             & 30       \\
AF               & 285           & 225          & 30             & 30       \\
IAVB             & 258           & 206          & 26             & 26       \\
STach, TAb       & 258           & 206          & 26             & 26       \\
TAb, QAb         & 226           & 178          & 24             & 24       \\
RBBB             & 224           & 176          & 24             & 24       \\
LQRSV            & 207           & 163          & 22             & 22       \\
IAVB, SB         & 203           & 159          & 22             & 22       \\
TAb, TInv        & 187           & 147          & 20             & 20       \\
TAb, SB          & 178           & 142          & 18             & 18       \\
PVC              & 134           & 106          & 14             & 14       \\
SA, SB           & 134           & 106          & 14             & 14       \\
IRBBB            & 128           & 100          & 14             & 14       \\
AF, TAb          & 126           & 98           & 14             & 14       \\
LAD, SB          & 125           & 97           & 14             & 14       \\
LAnFB            & 96            & 76           & 10             & 10       \\
\midrule
...\\
\midrule
LAD, STach, IAVB, BBB, LBBB, PAC & 2 & 0 & 0 & 2 \\
AF, TInv, IRBBB                  & 2 & 0 & 0 & 2 \\
PAC, SA, SB                      & 2 & 0 & 0 & 2 \\
QAb, TAb, SB, LQT, SA            & 2 & 0 & 0 & 2 \\
LAnFB, TAb, SA                   & 2 & 0 & 0 & 2 \\
IAVB, SA, PAC                    & 2 & 0 & 0 & 2 \\
NSIVCB, AFL, SB                  & 2 & 0 & 0 & 2 \\
LAnFB, LQT, SB                   & 2 & 0 & 0 & 2 \\
AF, TAb, PVC, QAb                & 2 & 0 & 0 & 2 \\
NSIVCB, LAD, LQT, SB             & 2 & 0 & 0 & 2 \\
AF, TAb, IRBBB, RBBB, LQT        & 2 & 0 & 0 & 2 \\
PAC, TAb, SA, QAb                & 2 & 0 & 0 & 2 \\
LAD, TAb, SB, LQT, PAC           & 2 & 0 & 0 & 2 \\
RAD, TAb, QAb                    & 2 & 0 & 0 & 2 \\
LAD, TAb, IRBBB, TInv, AFL       & 2 & 0 & 0 & 2 \\
LAD, TAb, TInv, STach            & 2 & 0 & 0 & 2 \\
LAD, TAb, IAVB, TInv, LQT, PAC   & 2 & 0 & 0 & 2 \\
LAD, STach, PVC, IRBBB           & 2 & 0 & 0 & 2 \\
RAD, SA, TInv                    & 2 & 0 & 0 & 2 \\
QAb, TInv, SB                    & 2 & 0 & 0 & 2 \\
AF, TAb, TInv, AFL               & 2 & 0 & 0 & 2 \\
AF, TAb, PAC                     & 2 & 0 & 0 & 2 \\
AF, TAb, IRBBB, RBBB, SB         & 2 & 0 & 0 & 2 \\
LAD, IAVB, BBB, LBBB, PVC        & 2 & 0 & 0 & 2 \\
BBB, RBBB, PAC                   & 1 & 0 & 0 & 1 \\
RAD, TAb, TInv, IRBBB            & 1 & 0 & 0 & 1 \\
\bottomrule
\end{tabular}
\end{table}

\clearpage

\begin{figure}[ht]
  \centering
  \includegraphics[width=\textwidth]{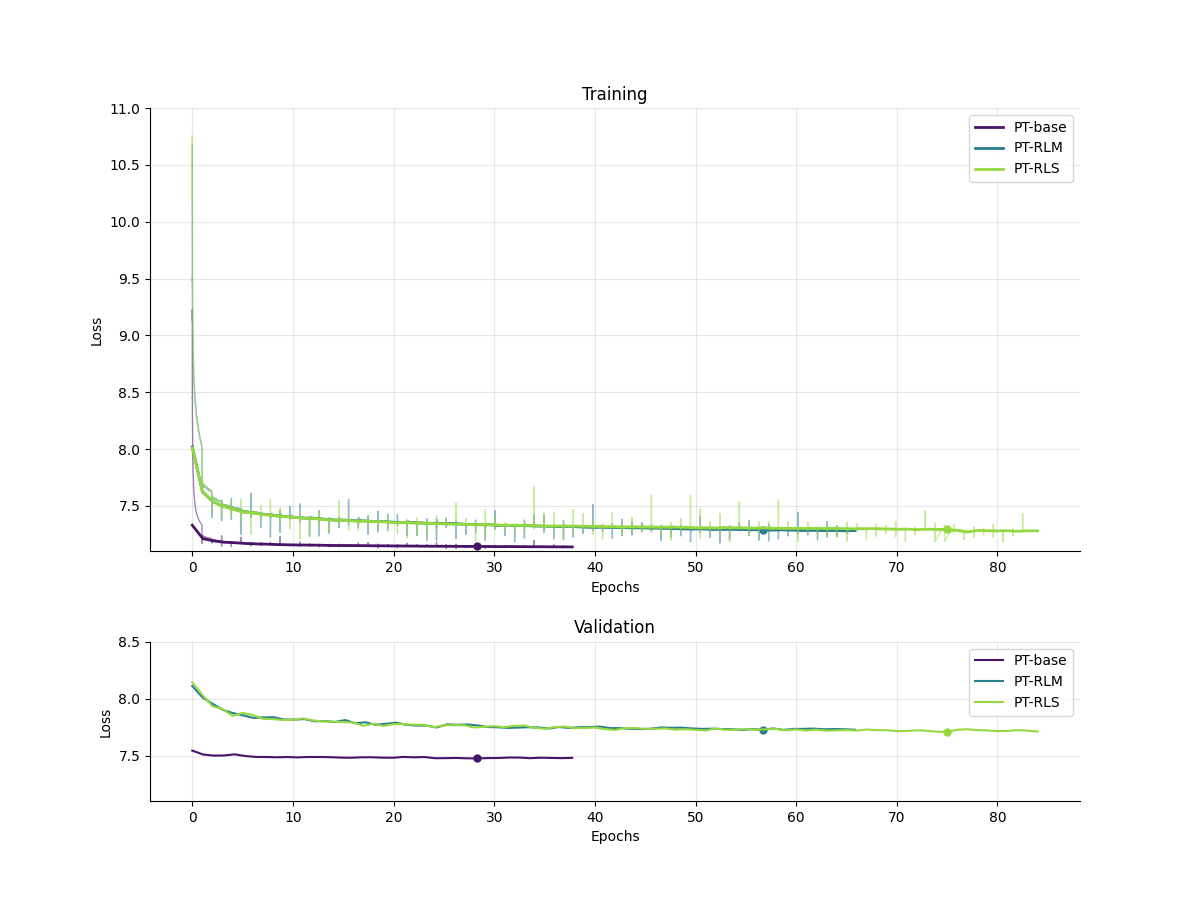}
  \caption{Training and validation loss curves during pre-training for the three models (PT-base, PT-RLM and PT-RLS).}
  \label{app_fig:loss_curve_pt}
\end{figure}

\begin{figure}[ht]
  \centering
  \includegraphics[width=\textwidth, trim=100 100 100 100, clip]{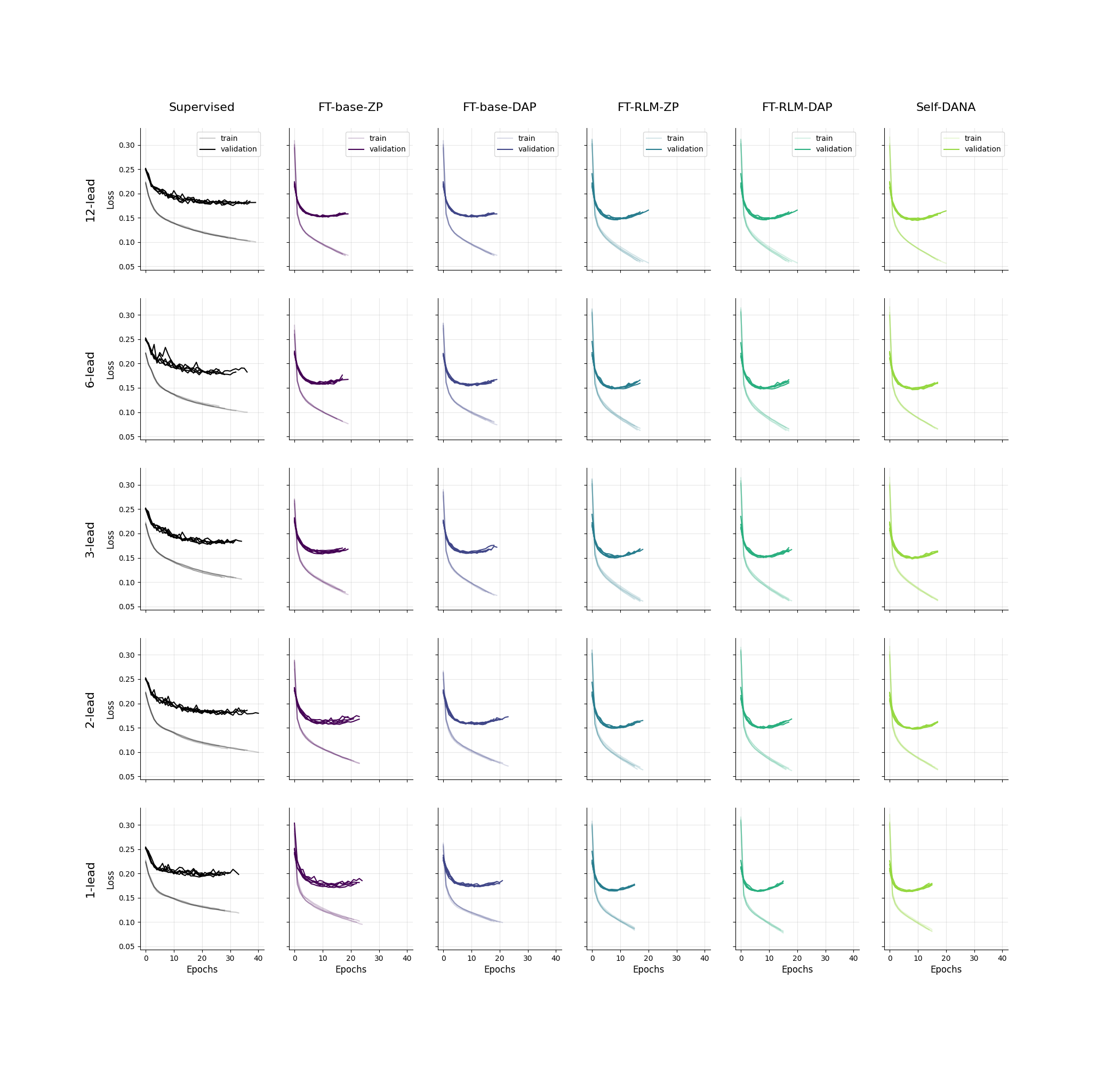}
  \caption{Training and validation loss curves during fine-tuning for all the models and reduced-lead configurations of our experiments.}
  \label{app_fig:loss_curve_ft}
\end{figure}

\paragraph{Learning curves}
Figure \ref{app_fig:loss_curve_pt} and \ref{app_fig:loss_curve_ft} report the validation and training loss obtained, respectively, during the pre-training and fine-tuning of all the models and configurations considered in our experiments.

\clearpage

\paragraph{Other single-lead models}
Table \ref{app_tab:single_lead_perf} reports the performance obtained by repeating our experiments with each of the 12 standard ECG leads in a single-lead setting. Self-DANA achieves the best performance for all 12 single-lead configurations, except for I, III, and aVF, for which it is comparable to the best results.

\begin{table}[ht]
\caption{CinC score obtained in experiments \textit{(i)}, \textit{(ii)} and \textit{(iii)} on the Georgia test sets of the 12 single-lead configurations. Results are reported as \textit{mean $\pm$ standard deviation}. Best results (highest CinC score) in bold.}
\label{app_tab:single_lead_perf}
\centering
\begin{tabular*}{\textwidth}{@{\extracolsep\fill}lcccccc}
\toprule
\multicolumn{1}{c}{} & \multicolumn{1}{c}{I} & \multicolumn{1}{c}{II} & \multicolumn{1}{c}{III} & \multicolumn{1}{c}{aVR} & \multicolumn{1}{c}{aVL} & \multicolumn{1}{c}{aVF}\\
\midrule
Supervised         & \begin{tabular}[c]{@{}c@{}}0.547\\±0.003\end{tabular}                         & \begin{tabular}[c]{@{}c@{}}0.563\\ ±0.003\end{tabular}          & \begin{tabular}[c]{@{}c@{}}0.510\\ ±0.008\end{tabular}          & \begin{tabular}[c]{@{}c@{}}0.569\\ ±0.006\end{tabular}          & \begin{tabular}[c]{@{}c@{}}0.516\\ ±0.003\end{tabular}          & \begin{tabular}[c]{@{}c@{}}0.536\\ ±0.003\end{tabular}          \\
\midrule
FT-base-ZP         & \begin{tabular}[c]{@{}c@{}}0.562\\ ±0.004\end{tabular}          & \begin{tabular}[c]{@{}c@{}}0.577\\ ±0.005\end{tabular}          & \begin{tabular}[c]{@{}c@{}}0.539\\ ±0.008\end{tabular}          & \begin{tabular}[c]{@{}c@{}}0.579\\ ±0.001\end{tabular}          & \begin{tabular}[c]{@{}c@{}}0.541\\ ±0.005\end{tabular}          & \begin{tabular}[c]{@{}c@{}}0.554\\ ±0.006\end{tabular}          \\
FT-base-DAP        & \begin{tabular}[c]{@{}c@{}}0.568\\ ±0.005\end{tabular}          & \begin{tabular}[c]{@{}c@{}}0.582\\ ±0.003\end{tabular}          & \begin{tabular}[c]{@{}c@{}}0.545\\ ±0.004\end{tabular}          & \begin{tabular}[c]{@{}c@{}}0.582\\ ±0.003\end{tabular}          & \begin{tabular}[c]{@{}c@{}}0.544\\ ±0.002\end{tabular}          & \begin{tabular}[c]{@{}c@{}}0.558\\ ±0.003\end{tabular}          \\
\midrule
FT-RLM-ZP          & \textbf{\begin{tabular}[c]{@{}c@{}}0.585\\ ±0.004\end{tabular}} & \begin{tabular}[c]{@{}c@{}}0.599\\ ±0.008\end{tabular}          & \begin{tabular}[c]{@{}c@{}}0.560\\ ±0.004\end{tabular}          & \begin{tabular}[c]{@{}c@{}}0.597\\ ±0.008\end{tabular}          & \begin{tabular}[c]{@{}c@{}}0.562\\ ±0.004\end{tabular}          & \textbf{\begin{tabular}[c]{@{}c@{}}0.575\\ ±0.005\end{tabular}} \\
FT-RLM-DAP         & \begin{tabular}[c]{@{}c@{}}0.578\\ ±0.003\end{tabular}          & \begin{tabular}[c]{@{}c@{}}0.596\\ ±0.010\end{tabular}          & \textbf{\begin{tabular}[c]{@{}c@{}}0.567\\ ±0.008\end{tabular}} & \begin{tabular}[c]{@{}c@{}}0.589\\ ±0.010\end{tabular}          & \begin{tabular}[c]{@{}c@{}}0.557\\ ±0.005\end{tabular}          & \textbf{\begin{tabular}[c]{@{}c@{}}0.575\\ ±0.005\end{tabular}} \\
\midrule
\textbf{Self-DANA} & \begin{tabular}[c]{@{}c@{}}0.583\\ ±0.004\end{tabular}          & \textbf{\begin{tabular}[c]{@{}c@{}}0.600\\ ±0.005\end{tabular}} & \begin{tabular}[c]{@{}c@{}}0.564\\ ±0.005\end{tabular}          & \textbf{\begin{tabular}[c]{@{}c@{}}0.607\\ ±0.008\end{tabular}} & \textbf{\begin{tabular}[c]{@{}c@{}}0.567\\ ±0.004\end{tabular}} & \begin{tabular}[c]{@{}c@{}}0.574\\ ±0.005\end{tabular}         
\\
\bottomrule
\\
\toprule
\multicolumn{1}{c}{} & \multicolumn{1}{c}{v1} & \multicolumn{1}{c}{v2} & \multicolumn{1}{c}{v3} & \multicolumn{1}{c}{v4} & \multicolumn{1}{c}{v5} & \multicolumn{1}{c}{v6}\\
\midrule
Supervised         & \begin{tabular}[c]{@{}c@{}}0.531\\ ±0.006\end{tabular}          & \begin{tabular}[c]{@{}c@{}}0.533\\ ±0.007\end{tabular}          & \begin{tabular}[c]{@{}c@{}}0.541\\ ±0.005\end{tabular}          & \begin{tabular}[c]{@{}c@{}}0.551\\ ±0.009\end{tabular}          & \begin{tabular}[c]{@{}c@{}}0.548\\ ±0.006\end{tabular}          & \begin{tabular}[c]{@{}c@{}}0.545\\ ±0.007\end{tabular}          \\
\midrule
FT-base-ZP         & \begin{tabular}[c]{@{}c@{}}0.549\\ ±0.004\end{tabular}          & \begin{tabular}[c]{@{}c@{}}0.550\\ ±0.002\end{tabular}          & \begin{tabular}[c]{@{}c@{}}0.559\\ ±0.003\end{tabular}          & \begin{tabular}[c]{@{}c@{}}0.567\\ ±0.007\end{tabular}          & \begin{tabular}[c]{@{}c@{}}0.569\\ ±0.004\end{tabular}          & \begin{tabular}[c]{@{}c@{}}0.572\\ ±0.003\end{tabular}          \\
FT-base-DAP        & \begin{tabular}[c]{@{}c@{}}0.550\\ ±0.002\end{tabular}          & \begin{tabular}[c]{@{}c@{}}0.550\\ ±0.002\end{tabular}          & \begin{tabular}[c]{@{}c@{}}0.560\\ ±0.008\end{tabular}          & \begin{tabular}[c]{@{}c@{}}0.572\\ ±0.004\end{tabular}          & \begin{tabular}[c]{@{}c@{}}0.579\\ ±0.002\end{tabular}          & \begin{tabular}[c]{@{}c@{}}0.575\\ ±0.002\end{tabular}          \\
\midrule
FT-RLM-ZP          & \begin{tabular}[c]{@{}c@{}}0.568\\ ±0.004\end{tabular}          & \textbf{\begin{tabular}[c]{@{}c@{}}0.573\\ ±0.003\end{tabular}} & \begin{tabular}[c]{@{}c@{}}0.578\\ ±0.009\end{tabular}          & \begin{tabular}[c]{@{}c@{}}0.589\\ ±0.010\end{tabular}          & \begin{tabular}[c]{@{}c@{}}0.590\\ ±0.006\end{tabular}          & \begin{tabular}[c]{@{}c@{}}0.585\\ ±0.006\end{tabular}          \\
FT-RLM-DAP         & \begin{tabular}[c]{@{}c@{}}0.571\\ ±0.001\end{tabular}          & \begin{tabular}[c]{@{}c@{}}0.569\\ ±0.003\end{tabular}          & \begin{tabular}[c]{@{}c@{}}0.573\\ ±0.006\end{tabular}          & \begin{tabular}[c]{@{}c@{}}0.579\\ ±0.007\end{tabular}          & \begin{tabular}[c]{@{}c@{}}0.590\\ ±0.002\end{tabular}          & \begin{tabular}[c]{@{}c@{}}0.584\\ ±0.006\end{tabular}          \\
\midrule
\textbf{Self-DANA} & \textbf{\begin{tabular}[c]{@{}c@{}}0.572\\ ±0.011\end{tabular}} & \textbf{\begin{tabular}[c]{@{}c@{}}0.573\\ ±0.004\end{tabular}} & \textbf{\begin{tabular}[c]{@{}c@{}}0.584\\ ±0.005\end{tabular}} & \textbf{\begin{tabular}[c]{@{}c@{}}0.596\\ ±0.005\end{tabular}} & \textbf{\begin{tabular}[c]{@{}c@{}}0.597\\ ±0.004\end{tabular}} & \textbf{\begin{tabular}[c]{@{}c@{}}0.595\\ ±0.003\end{tabular}}\\
\bottomrule
\end{tabular*}
\end{table}



\newpage



\clearpage
\bibliography{bibliography}

\begin{thebibliography}{39}
\providecommand{\natexlab}[1]{#1}
\providecommand{\url}[1]{\texttt{#1}}
\expandafter\ifx\csname urlstyle\endcsname\relax
  \providecommand{\doi}[1]{doi: #1}\else
  \providecommand{\doi}{doi: \begingroup \urlstyle{rm}\Url}\fi

\bibitem[Bommasani et~al.(2022)Bommasani, Hudson, Adeli, Altman, Arora, von Arx, Bernstein, Bohg, Bosselut, Brunskill, Brynjolfsson, Buch, Card, Castellon, Chatterji, Chen, Creel, Davis, Demszky, Donahue, Doumbouya, Durmus, Ermon, Etchemendy, Ethayarajh, Fei-Fei, Finn, Gale, Gillespie, Goel, Goodman, Grossman, Guha, Hashimoto, Henderson, Hewitt, Ho, Hong, Hsu, Huang, Icard, Jain, Jurafsky, Kalluri, Karamcheti, Keeling, Khani, Khattab, Koh, Krass, Krishna, Kuditipudi, Kumar, Ladhak, Lee, Lee, Leskovec, Levent, Li, Li, Ma, Malik, Manning, Mirchandani, Mitchell, Munyikwa, Nair, Narayan, Narayanan, Newman, Nie, Niebles, Nilforoshan, Nyarko, Ogut, Orr, Papadimitriou, Park, Piech, Portelance, Potts, Raghunathan, Reich, Ren, Rong, Roohani, Ruiz, Ryan, Ré, Sadigh, Sagawa, Santhanam, Shih, Srinivasan, Tamkin, Taori, Thomas, Tramèr, Wang, Wang, Wu, Wu, Wu, Xie, Yasunaga, You, Zaharia, Zhang, Zhang, Zhang, Zhang, Zheng, Zhou, and Liang]{Bommasani2021_FMdef}
Rishi Bommasani, Drew~A. Hudson, Ehsan Adeli, Russ Altman, Simran Arora, Sydney von Arx, Michael~S. Bernstein, Jeannette Bohg, Antoine Bosselut, Emma Brunskill, Erik Brynjolfsson, Shyamal Buch, Dallas Card, Rodrigo Castellon, Niladri Chatterji, Annie Chen, Kathleen Creel, Jared~Quincy Davis, Dora Demszky, Chris Donahue, Moussa Doumbouya, Esin Durmus, Stefano Ermon, John Etchemendy, Kawin Ethayarajh, Li~Fei-Fei, Chelsea Finn, Trevor Gale, Lauren Gillespie, Karan Goel, Noah Goodman, Shelby Grossman, Neel Guha, Tatsunori Hashimoto, Peter Henderson, John Hewitt, Daniel~E. Ho, Jenny Hong, Kyle Hsu, Jing Huang, Thomas Icard, Saahil Jain, Dan Jurafsky, Pratyusha Kalluri, Siddharth Karamcheti, Geoff Keeling, Fereshte Khani, Omar Khattab, Pang~Wei Koh, Mark Krass, Ranjay Krishna, Rohith Kuditipudi, Ananya Kumar, Faisal Ladhak, Mina Lee, Tony Lee, Jure Leskovec, Isabelle Levent, Xiang~Lisa Li, Xuechen Li, Tengyu Ma, Ali Malik, Christopher~D. Manning, Suvir Mirchandani, Eric Mitchell, Zanele Munyikwa, Suraj Nair,
  Avanika Narayan, Deepak Narayanan, Ben Newman, Allen Nie, Juan~Carlos Niebles, Hamed Nilforoshan, Julian Nyarko, Giray Ogut, Laurel Orr, Isabel Papadimitriou, Joon~Sung Park, Chris Piech, Eva Portelance, Christopher Potts, Aditi Raghunathan, Rob Reich, Hongyu Ren, Frieda Rong, Yusuf Roohani, Camilo Ruiz, Jack Ryan, Christopher Ré, Dorsa Sadigh, Shiori Sagawa, Keshav Santhanam, Andy Shih, Krishnan Srinivasan, Alex Tamkin, Rohan Taori, Armin~W. Thomas, Florian Tramèr, Rose~E. Wang, William Wang, Bohan Wu, Jiajun Wu, Yuhuai Wu, Sang~Michael Xie, Michihiro Yasunaga, Jiaxuan You, Matei Zaharia, Michael Zhang, Tianyi Zhang, Xikun Zhang, Yuhui Zhang, Lucia Zheng, Kaitlyn Zhou, and Percy Liang.
\newblock On the opportunities and risks of foundation models, 2022.
\newblock URL \url{https://arxiv.org/abs/2108.07258}.

\bibitem[Devlin et~al.(2019)Devlin, Chang, Lee, Google, and Language]{Devlin2019_BERT}
Jacob Devlin, Ming-Wei Chang, Kenton Lee, Kristina~Toutanova Google, and A~I Language.
\newblock Bert: Pre-training of deep bidirectional transformers for language understanding.
\newblock \emph{Proceedings of the 2019 Conference of the North}, pages 4171--4186, 2019.
\newblock \doi{10.18653/V1/N19-1423}.
\newblock URL \url{https://aclanthology.org/N19-1423/}.

\bibitem[Brown et~al.(2020)Brown, Mann, Ryder, Subbiah, Kaplan, Dhariwal, Neelakantan, Shyam, Sastry, Askell, Agarwal, Herbert-Voss, Krueger, Henighan, Child, Ramesh, Ziegler, Wu, Winter, Hesse, Chen, Sigler, Litwin, Gray, Chess, Clark, Berner, McCandlish, Radford, Sutskever, and Amodei]{Brown2020_GPT}
Tom~B. Brown, Benjamin Mann, Nick Ryder, Melanie Subbiah, Jared Kaplan, Prafulla Dhariwal, Arvind Neelakantan, Pranav Shyam, Girish Sastry, Amanda Askell, Sandhini Agarwal, Ariel Herbert-Voss, Gretchen Krueger, Tom Henighan, Rewon Child, Aditya Ramesh, Daniel~M. Ziegler, Jeffrey Wu, Clemens Winter, Christopher Hesse, Mark Chen, Eric Sigler, Mateusz Litwin, Scott Gray, Benjamin Chess, Jack Clark, Christopher Berner, Sam McCandlish, Alec Radford, Ilya Sutskever, and Dario Amodei.
\newblock Language models are few-shot learners.
\newblock \emph{Advances in Neural Information Processing Systems}, 2020-December, 5 2020.
\newblock ISSN 10495258.
\newblock URL \url{https://arxiv.org/pdf/2005.14165}.

\bibitem[Radford et~al.(2021)Radford, Kim, Hallacy, Ramesh, Goh, Agarwal, Sastry, Askell, Mishkin, Clark, Krueger, and Sutskever]{Radford2021_CLIP}
Alec Radford, Jong~Wook Kim, Chris Hallacy, Aditya Ramesh, Gabriel Goh, Sandhini Agarwal, Girish Sastry, Amanda Askell, Pamela Mishkin, Jack Clark, Gretchen Krueger, and Ilya Sutskever.
\newblock Learning transferable visual models from natural language supervision.
\newblock \emph{Proceedings of Machine Learning Research}, 139:\penalty0 8748--8763, 2 2021.
\newblock ISSN 26403498.
\newblock URL \url{https://arxiv.org/pdf/2103.00020}.

\bibitem[Baevski et~al.(2020)Baevski, Zhou, Mohamed, and Auli]{Baevski2020_wav2vec2}
Alexei Baevski, Henry Zhou, Abdelrahman Mohamed, and Michael Auli.
\newblock wav2vec 2.0: A framework for self-supervised learning of speech representations.
\newblock \emph{Advances in Neural Information Processing Systems}, 2020-December, 6 2020.
\newblock ISSN 10495258.
\newblock URL \url{https://arxiv.org/pdf/2006.11477}.

\bibitem[Hsu et~al.(2021)Hsu, Bolte, Tsai, Lakhotia, Salakhutdinov, and Mohamed]{Hsu2021_HuBERT}
Wei~Ning Hsu, Benjamin Bolte, Yao Hung~Hubert Tsai, Kushal Lakhotia, Ruslan Salakhutdinov, and Abdelrahman Mohamed.
\newblock Hubert: Self-supervised speech representation learning by masked prediction of hidden units.
\newblock \emph{IEEE/ACM Transactions on Audio Speech and Language Processing}, 29:\penalty0 3451--3460, 6 2021.
\newblock ISSN 23299304.
\newblock \doi{10.1109/TASLP.2021.3122291}.
\newblock URL \url{https://arxiv.org/pdf/2106.07447}.

\bibitem[Thapa et~al.(2024)Thapa, He, Kjær, Moore, Ganjoo, Mignot, and Zou]{Thapa2025_sleepFM}
Rahul Thapa, Bryan He, Magnus~Ruud Kjær, Hyatt Moore, Gauri Ganjoo, Emmanuel Mignot, and James Zou.
\newblock Sleepfm: Multi-modal representation learning for sleep across brain activity, ecg and respiratory signals.
\newblock \emph{Proceedings of Machine Learning Research}, 235:\penalty0 48019--48037, 5 2024.
\newblock ISSN 26403498.
\newblock URL \url{https://arxiv.org/pdf/2405.17766}.

\bibitem[Vaid et~al.(2023)Vaid, Jiang, Sawant, Lerakis, Argulian, Ahuja, Lampert, Charney, Greenspan, Narula, Glicksberg, and Nadkarni]{Vaid2023_HeartBEiT}
Akhil Vaid, Joy Jiang, Ashwin Sawant, Stamatios Lerakis, Edgar Argulian, Yuri Ahuja, Joshua Lampert, Alexander Charney, Hayit Greenspan, Jagat Narula, Benjamin Glicksberg, and Girish~N. Nadkarni.
\newblock A foundational vision transformer improves diagnostic performance for electrocardiograms.
\newblock \emph{npj Digital Medicine}, 6:\penalty0 1--8, 12 2023.
\newblock ISSN 23986352.
\newblock \doi{10.1038/S41746-023-00840-9}.
\newblock URL \url{https://www.nature.com/articles/s41746-023-00840-9}.

\bibitem[Pup and Atzori(2023)]{DelPup2023_review_SSL_biosig}
Federico~Del Pup and Manfredo Atzori.
\newblock Applications of self-supervised learning to biomedical signals: A survey.
\newblock \emph{IEEE Access}, 11:\penalty0 144180--144203, 2023.
\newblock ISSN 21693536.
\newblock \doi{10.1109/ACCESS.2023.3344531}.

\bibitem[Han et~al.(2024)Han, Liu, Zhang, and Ding]{Han2024_reviewFM}
Yu~Han, Xiaofeng Liu, Xiang Zhang, and Cheng Ding.
\newblock Foundation models in electrocardiogram: A review, 2024.
\newblock URL \url{https://arxiv.org/abs/2410.19877}.

\bibitem[McKeen et~al.(2024)McKeen, Oliva, Masood, Toma, Rubin, and Wang]{McKeen2024_ECGFM}
Kaden McKeen, Laura Oliva, Sameer Masood, Augustin Toma, Barry Rubin, and Bo~Wang.
\newblock Ecg-fm: An open electrocardiogram foundation model, 2024.
\newblock URL \url{https://arxiv.org/abs/2408.05178}.

\bibitem[Abbaspourazad et~al.(2023)Abbaspourazad, Elachqar, Miller, Emrani, Nallasamy, and Shapiro]{Abbaspourazad2023}
Salar Abbaspourazad, Oussama Elachqar, Andrew~C. Miller, Saba Emrani, Udhyakumar Nallasamy, and Ian Shapiro.
\newblock Large-scale training of foundation models for wearable biosignals.
\newblock \emph{12th International Conference on Learning Representations, ICLR 2024}, 12 2023.
\newblock URL \url{https://arxiv.org/pdf/2312.05409}.

\bibitem[Yu et~al.(2024)Yu, Guo, and Sano]{Yu2024}
Han Yu, Peikun Guo, and Akane Sano.
\newblock Ecg semantic integrator (esi): A foundation ecg model pretrained with llm-enhanced cardiological text, 2024.
\newblock URL \url{https://arxiv.org/abs/2405.19366}.

\bibitem[Mathew et~al.(2024)Mathew, Barbosa, Prince, and Venkatraman]{Mathew2024}
George Mathew, Daniel Barbosa, John Prince, and Subramaniam Venkatraman.
\newblock Foundation models for cardiovascular disease detection via biosignals from digital stethoscopes.
\newblock \emph{npj Cardiovascular Health 2024 1:1}, 1:\penalty0 1--13, 10 2024.
\newblock ISSN 2948-2836.
\newblock \doi{10.1038/s44325-024-00027-5}.
\newblock URL \url{https://www.nature.com/articles/s44325-024-00027-5}.

\bibitem[Bouzid et~al.(2022)Bouzid, Al-Zaiti, Bond, and Sejdi{\'c}]{bouzid2022remote}
Zeineb Bouzid, Salah~S Al-Zaiti, Raymond Bond, and Ervin Sejdi{\'c}.
\newblock Remote and wearable ecg devices with diagnostic abilities in adults: a state-of-the-science scoping review.
\newblock \emph{Heart Rhythm}, 19\penalty0 (7):\penalty0 1192--1201, 2022.

\bibitem[Gopal et~al.(2021)Gopal, Han, Raghupathi, Ng, Tison, and Rajpurkar]{Gopal2021_3KG}
Bryan Gopal, Ryan Han, Gautham Raghupathi, Andrew Ng, Geoff Tison, and Pranav Rajpurkar.
\newblock 3kg: Contrastive learning of 12-lead electrocardiograms using physiologically-inspired augmentations.
\newblock \emph{Proceedings of Machine Learning Research}, 158:\penalty0 156--167, 4 2021.
\newblock ISSN 26403498.
\newblock URL \url{https://arxiv.org/pdf/2106.04452}.

\bibitem[Liu et~al.(2023)Liu, Li, Zhang, Chang, Wang, He, and Huang]{Liu2023_JCDCL}
Wenhan Liu, Zhoutong Li, Huaicheng Zhang, Sheng Chang, Hao Wang, Jin He, and Qijun Huang.
\newblock Dense lead contrast for self-supervised representation learning of multilead electrocardiograms.
\newblock \emph{Information Sciences}, 634:\penalty0 189--205, 7 2023.
\newblock ISSN 0020-0255.
\newblock \doi{10.1016/J.INS.2023.03.099}.
\newblock URL \url{https://www.sciencedirect.com/science/article/pii/S002002552300422X}.

\bibitem[Oh et~al.(2022)Oh, Chung, Hong, and Choi]{Oh2022_RLM}
Jungwoo Oh, Hyunseung Chung, Dong-Gyun Hong, and Edward Choi.
\newblock Lead-agnostic self-supervised learning for local and global representations of electrocardiogram.
\newblock \emph{Proceedings of Machine Learning Research}, 174:\penalty0 2022, 2022.
\newblock URL \url{https://github.com/Jwoo5/fairseq-signals}.

\bibitem[Malekzadeh et~al.(2020)Malekzadeh, Clegg, Cavallaro, and Haddadi]{Malekzadeh2020_DANA}
Mohammad Malekzadeh, Richard~G. Clegg, Andrea Cavallaro, and Hamed Haddadi.
\newblock Dana: Dimension-adaptive neural architecture for multivariate sensor data.
\newblock \emph{Proceedings of the ACM on Interactive, Mobile, Wearable and Ubiquitous Technologies}, 5:\penalty0 120, 8 2020.
\newblock \doi{10.1145/3478074}.
\newblock URL \url{http://dx.doi.org/10.1145/3478074}.

\bibitem[Li et~al.(2025)Li, Aguirre, Moura, Liu, Zhong, Sun, Clifford, Westover, and Hong]{Li2024_ECGFounder}
Jun Li, Aaron Aguirre, Junior Moura, Che Liu, Lanhai Zhong, Chenxi Sun, Gari Clifford, Brandon Westover, and Shenda Hong.
\newblock An electrocardiogram foundation model built on over 10 million recordings with external evaluation across multiple domains, 2025.
\newblock URL \url{https://arxiv.org/abs/2410.04133}.

\bibitem[Yang et~al.(2023)Yang, Westover, and Sun]{Yang2023_BIOT}
Chaoqi Yang, M~Brandon Westover, and Jimeng Sun.
\newblock Biot: Biosignal transformer for cross-data learning in the wild.
\newblock \emph{Advances in Neural Information Processing Systems}, 36:\penalty0 78240--78260, 12 2023.

\bibitem[Reyna et~al.(2021)Reyna, Sadr, Alday, Gu, Shah, Robichaux, Rad, Elola, Seyedi, Ansari, Ghanbari, Li, Sharma, and Clifford]{cinc21_1}
Matthew~A Reyna, Nadi Sadr, Erick A~Perez Alday, Annie Gu, Amit~J Shah, Chad Robichaux, Ali~Bahrami Rad, Andoni Elola, Salman Seyedi, Sardar Ansari, Hamid Ghanbari, Qiao Li, Ashish Sharma, and Gari~D Clifford.
\newblock Will two do? varying dimensions in electrocardiography: The physionet/computing in cardiology challenge 2021.
\newblock In \emph{Computing in Cardiology}, volume~48, pages 1--4, 2021.

\bibitem[Reyna et~al.(2022{\natexlab{a}})Reyna, Sadr, Alday, Gu, Shah, Robichaux, Rad, Elola, Seyedi, Ansari, Ghanbari, Li, Sharma, and Clifford]{cinc21_2}
Matthew~A Reyna, Nadi Sadr, Erick A~Perez Alday, Annie Gu, Amit~J Shah, Chad Robichaux, Ali~Bahrami Rad, Andoni Elola, Salman Seyedi, Sardar Ansari, Hamid Ghanbari, Qiao Li, Ashish Sharma, and Gari~D Clifford.
\newblock Issues in the automated classification of multilead ecgs using heterogeneous labels and populations.
\newblock \emph{Physiological Measurement}, 2022{\natexlab{a}}.

\bibitem[Chen et~al.(2020)Chen, Kornblith, Norouzi, and Hinton]{Chen2020_simclr}
Ting Chen, Simon Kornblith, Mohammad Norouzi, and Geoffrey Hinton.
\newblock A simple framework for contrastive learning of visual representations.
\newblock \emph{37th International Conference on Machine Learning, ICML 2020}, PartF168147-3:\penalty0 1575--1585, 2 2020.
\newblock URL \url{https://arxiv.org/pdf/2002.05709}.

\bibitem[Ribeiro et~al.(2020)Ribeiro, Ribeiro, Paixão, Oliveira, Gomes, Canazart, Ferreira, Andersson, Macfarlane, Wagner, Schön, and Ribeiro]{code_train_paper}
Antônio~H. Ribeiro, Manoel~Horta Ribeiro, Gabriela~M.M. Paixão, Derick~M. Oliveira, Paulo~R. Gomes, Jéssica~A. Canazart, Milton~P.S. Ferreira, Carl~R. Andersson, Peter~W. Macfarlane, Meira Wagner, Thomas~B. Schön, and Antonio Luiz~P. Ribeiro.
\newblock Automatic diagnosis of the 12-lead ecg using a deep neural network.
\newblock \emph{Nature Communications}, 11:\penalty0 1--9, 12 2020.
\newblock ISSN 20411723.
\newblock \doi{10.1038/S41467-020-15432-4}.
\newblock URL \url{https://www.nature.com/articles/s41467-020-15432-4}.

\bibitem[Ribeiro et~al.(2021)Ribeiro, Paixao, Lima, Ribeiro, Filho, Gomes, Oliveira, Jr, Schon, and Ribeiro]{code_training_data}
Antônio~H. Ribeiro, Gabriela~M.M. Paixao, Emilly~M. Lima, Manoel~Horta Ribeiro, Marcelo M.~Pinto Filho, Paulo~R. Gomes, Derick~M. Oliveira, Wagner~Meira Jr, Thömas~B Schon, and Antonio Luiz~P. Ribeiro.
\newblock Code-15
\newblock URL \url{https://zenodo.org/records/4916206}.
\newblock v1.0.0; Accessed: 2024-01-17; CC BY 4.0 license.

\bibitem[Lai et~al.(2023)Lai, Tan, Wang, Ji, Guo, Han, Shi, Feng, and Yang]{off_test_paper}
Jiewei Lai, Huixin Tan, Jinliang Wang, Lei Ji, Jun Guo, Baoshi Han, Yajun Shi, Qianjin Feng, and Wei Yang.
\newblock Practical intelligent diagnostic algorithm for wearable 12-lead ecg via self-supervised learning on large-scale dataset.
\newblock \emph{Nature Communications 2023 14:1}, 14:\penalty0 1--13, 6 2023.
\newblock ISSN 2041-1723.
\newblock \doi{10.1038/s41467-023-39472-8}.
\newblock URL \url{https://www.nature.com/articles/s41467-023-39472-8}.

\bibitem[off(2023)]{off_test_data}
Offline test set of ecg multi-label classfication, 2023.
\newblock URL \url{https://www.scidb.cn/en/detail?dataSetId=58c4a92d5a01414390a78160d335380d}.
\newblock v1; Accessed: 2024-05-02; MIT license.

\bibitem[Goldberger et~al.(2000 (June 13))Goldberger, Amaral, Glass, Hausdorff, Ivanov, Mark, Mietus, Moody, Peng, and Stanley]{PhysioNet}
A.~L. Goldberger, L.~A.~N. Amaral, L.~Glass, J.~M. Hausdorff, P.~Ch. Ivanov, R.~G. Mark, J.~E. Mietus, G.~B. Moody, C.-K. Peng, and H.~E. Stanley.
\newblock {PhysioBank, PhysioToolkit, and PhysioNet}: Components of a new research resource for complex physiologic signals.
\newblock \emph{Circulation}, 101\penalty0 (23):\penalty0 e215--e220, 2000 (June 13).
\newblock Circulation Electronic Pages: http://circ.ahajournals.org/content/101/23/e215.full PMID:1085218; doi: 10.1161/01.CIR.101.23.e215.

\bibitem[Reyna et~al.(2022{\natexlab{b}})Reyna, Sadr, Gu, Alday, Liu, Seyedi, Shah, and Clifford]{cinc21_data}
Matthew~A Reyna, Nadi Sadr, Annie Gu, Erick A~Perez Alday, Chengyu Liu, Salman Seyedi, Amit~J Shah, and Gari~D Clifford.
\newblock Will two do? varying dimensions in electrocardiography: The physionet/computing in cardiology challenge 2021.
\newblock PhysioNet, 2022{\natexlab{b}}.
\newblock URL \url{https://physionet.org/content/challenge-2021/1.0.3/}.
\newblock v1.0.3; Accessed: 2024-02-19; CC BY 4.0 license.

\bibitem[Zheng et~al.(2020{\natexlab{a}})Zheng, Chu, Struppa, Zhang, Yacoub, El-Askary, Chang, Ehwerhemuepha, Abudayyeh, Barrett, Fu, Yao, Li, Guo, and Rakovski]{ningbo_paper}
Jianwei Zheng, Huimin Chu, Daniele Struppa, Jianming Zhang, Magdi Yacoub, Hesham El-Askary, Anthony Chang, Louis Ehwerhemuepha, Islam Abudayyeh, Alexander Barrett, Guohua Fu, Hai Yao, Dongbo Li, Hangyuan Guo, and Cyril Rakovski.
\newblock Optimal multi-stage arrhythmia classification approach.
\newblock \emph{Scientific Reports}, 10, 02 2020{\natexlab{a}}.
\newblock \doi{10.1038/s41598-020-59821-7}.

\bibitem[Zheng et~al.(2020{\natexlab{b}})Zheng, Zhang, Danioko, Yao, Guo, and Rakovski]{chapman-shaoxing_paper}
Jianwei Zheng, Jianming Zhang, Sidy Danioko, Hai Yao, Hangyuan Guo, and Cyril Rakovski.
\newblock A 12-lead electrocardiogram database for arrhythmia research covering more than 10,000 patients.
\newblock \emph{Scientific Data}, 7, 02 2020{\natexlab{b}}.
\newblock \doi{10.1038/s41597-020-0386-x}.

\bibitem[Liu et~al.(2018)Liu, Liu, Zhao, Zhang, Wu, Xu, Liu, Ma, Wei, He, Li, Ng, and Kwee]{cpsc_paper}
Feifei Liu, Chengyu Liu, Lina Zhao, Xiangyu Zhang, Xiaoling Wu, Xiaoyan Xu, Yulin Liu, Caiyun Ma, Shoushui Wei, Zhiqiang He, Jianqing Li, Eddie Ng, and Yin Kwee.
\newblock An open access database for evaluating the algorithms of electrocardiogram rhythm and morphology abnormality detection.
\newblock \emph{Journal of Medical Imaging and Health Informatics}, 8:\penalty0 1368--1373, 2018.
\newblock \doi{10.1166/jmihi.2018.2442}.
\newblock URL \url{http://www.icbeb.org/Challenge.html}.

\bibitem[Nejedly et~al.(2021)Nejedly, Ivora, Smisek, Viscor, Koscova, Jurak, and Plesinger]{Nejedly2021_cinc21_winner}
Petr Nejedly, Adam Ivora, Radovan Smisek, Ivo Viscor, Zuzana Koscova, Pavel Jurak, and Filip Plesinger.
\newblock Classification of ecg using ensemble of residual cnns with attention mechanism.
\newblock \emph{Computing in Cardiology}, 48, 2021.
\newblock \doi{10.22489/CinC.2021.014}.

\bibitem[Soltanieh et~al.(2022)Soltanieh, Etemad, and Hashemi]{Soltanieh2022_aug}
Sahar Soltanieh, Ali Etemad, and Javad Hashemi.
\newblock Analysis of augmentations for contrastive ecg representation learning.
\newblock In \emph{2022 International Joint Conference on Neural Networks (IJCNN)}, pages 1--10, 2022.
\newblock \doi{10.1109/IJCNN55064.2022.9892600}.

\bibitem[Mehari and Strodthoff(2022)]{Mehari2022_SSLECG}
Temesgen Mehari and Nils Strodthoff.
\newblock Self-supervised representation learning from 12-lead ecg data.
\newblock \emph{Computers in Biology and Medicine}, 141:\penalty0 105114, 2 2022.
\newblock ISSN 0010-4825.
\newblock \doi{10.1016/J.COMPBIOMED.2021.105114}.
\newblock URL \url{https://www.sciencedirect.com/science/article/pii/S0010482521009082}.

\bibitem[Del~Pup et~al.(2024{\natexlab{a}})Del~Pup, Zanola, Tshimanga, Mazzon, and Atzori]{selfeeg_code}
Federico Del~Pup, Andrea Zanola, Louis~Fabrice Tshimanga, Paolo~Emilio Mazzon, and Manfredo Atzori.
\newblock selfeeg.
\newblock GitHub repository, 2024{\natexlab{a}}.
\newblock URL \url{https://github.com/MedMaxLab/selfEEG}.
\newblock v0.2.0; Accessed: 2024-10-09; MIT license.

\bibitem[Del~Pup et~al.(2024{\natexlab{b}})Del~Pup, Zanola, Tshimanga, Mazzon, and Atzori]{selfeeg_paper}
Federico Del~Pup, Andrea Zanola, Louis~Fabrice Tshimanga, Paolo~Emilio Mazzon, and Manfredo Atzori.
\newblock Selfeeg: A python library for self-supervised learning in electroencephalography.
\newblock \emph{Journal of Open Source Software}, 9\penalty0 (95):\penalty0 6224, 2024{\natexlab{b}}.
\newblock \doi{10.21105/joss.06224}.
\newblock URL \url{https://doi.org/10.21105/joss.06224}.

\bibitem[fai(2023)]{fairseq-signals_code}
fairseq-signals.
\newblock GitHub repository, 2023.
\newblock URL \url{https://github.com/Jwoo5/fairseq-signals}.
\newblock Accessed: 2024-10-09; MIT license.

\end{thebibliography}

\end{document}